\documentclass[letterpaper,10pt,nofootinbib,aps,tightenlines]{revtex4}
\usepackage{amsmath,amsfonts,amssymb}
\usepackage{mathrsfs}
\usepackage{graphicx}
\usepackage[english]{babel}

\newcommand{\be}{\begin{equation}}
\newcommand{\ee}{\end{equation}}
\newcommand{\bes}{\begin{equation*}}
\newcommand{\ees}{\end{equation*}}
\newcommand{\mn}{{\mu \nu}}
 
\newcommand{\nx}{{n_\chi}}

\usepackage{color}

\begin{document}

\title{Axion Gauge Field Inflation and Gravitational Leptogenesis:\\ A Lower Bound on B Modes from the Matter-Antimatter Asymmetry of the Universe}
\author{R. R. Caldwell}
\author{C. Devulder}
\affiliation{Department of Physics \& Astronomy, Dartmouth College, 6127 Wilder Laboratory, Hanover, NH 03755 USA}

\date{\today}

\begin{abstract}
We present a toy model of an axion gauge field inflation scenario that yields viable density and gravitational wave spectra. The scenario consists of an axionic inflaton in a steep potential that is effectively flattened by a coupling to a collection of non-Abelian gauge fields. The model predicts a blue-tilted gravitational wave spectrum that is dominated by one circular polarization, resulting in unique observational targets for cosmic microwave background and gravitational wave experiments. The handedness of the gravitational wave spectrum is incorporated in a model of leptogenesis through the axial-gravitational anomaly; assuming electroweak sphaeleron processes convert the lepton asymmetry into baryons, we predict an approximate lower bound on the tensor-to-scalar ratio $r \sim 3-4\times 10^{-2}$ for models that also explain the matter-antimatter asymmetry of the Universe. 
\end{abstract}

\maketitle

\section{Introduction}

Inflation is the leading paradigm for the hot Big Bang origin of the Universe \cite{Guth:1980zm,Linde:1981mu,Albrecht:1982wi}. The basic features of the inflationary scenario, notably a spatially flat Universe with a spectrum of nearly scale-invariant, adiabatic, gaussian-distributed density perturbations, are consistent with the growing catalog of experimental and observational data \cite{Ade:2015lrj}. Many models of inflation also predict a spectrum of primordial gravitational waves \cite{Starobinsky:1979ty}, which would leave a distinct imprint on the polarization pattern of the cosmic microwave background (CMB) \cite{Kamionkowski:1996ks,Zaldarriaga:1996xe,Kamionkowski:2015yta}. This ``B-mode" pattern is being actively pursued by a number of CMB experiments \cite{Fraisse:2011xz,Benson:2014qhw,Ahmed:2014ixy,Essinger-Hileman:2014pja,Ade:2015rva,Suzuki:2015zzg,Thornton:2016wjq,Stebor:2016hgt,Matsumura:2013aja,deBernardis:2017ofr}. Yet for all the successes, models of inflation that predict a potentially measurable level of gravitational waves typically require masses and field excursions that exceed the Planck scale, raising a spectre of instability against quantum gravitational corrections \cite{Lyth:1996im}. 

Chromo-natural inflation was proposed, in this context, as a new method for inflating with sub-Planckian masses \cite{Adshead:2012kp}. The model consists of an axionic inflaton with a shift symmetry, as found in natural inflation \cite{Freese:1990rb}
\be
V (\chi) = m^4 (1-\cos \chi/f)
\label{eqn:Vorig}
\ee
where $\chi$ is the scalar field and $m,\,f \ll M_P$ are the mass parameters. Alone, this would make the potential too steep to slow roll. However, the scenario features a coupling to a collection of non-Abelian gauge fields with a vacuum expectation value. In the simplest realization, the model posits that an SU(2) subgroup of an SU(N) is in a flavor-space locked configuration, whereby the global part of the SU(2) is identified with the O(3) rotational symmetry of spacetime. The exchange of energy introduced by the axion - Chern-Simons coupling serves to flatten the effective potential and bring about slow roll inflation.

The fluctuation spectra in chromo-natural inflation do not resemble the predictions in the simple case of single field inflation. In fact, the standard relationships that link the Hubble scale $H$, slow roll parameters $\epsilon_H = -\dot H/H^2$, $\eta_H = \epsilon_H - \ddot H/\dot H H$ with the spectrum amplitude $\Delta_\zeta^2$, scalar spectral index $n_s$, and tensor-to-scalar ratio $r$, do not apply. (See Ref.~\cite{Baumann:2009ds} for a review.) Instead, numerical calculations, supported by analytic approximations, reveal a red-tilted spectrum of density perturbations and a strongly amplified, chirally-asymmetric, blue-tilted spectrum of gravitational waves. In the original formulation of chromo-natural inflation, there is no satisfactory compromise between the predicted values of $n_s$ and $r$ that is consistent with observational constraints. Chromo-natural inflation and its variant, gauge-flation, as originally proposed, are thus ruled out \cite{Adshead:2013qp,Adshead:2013nka,Namba:2013kia}. 

In this work, we propose a toy model variation of the original model, which now satisfies current bounds on scalar and tensor spectra. Our clue comes from the tendency of the original model to produce a smaller $n_s$ and larger $r$ than would be expected based on the slow roll parameters. By modifying the axion potential into a form that naively predicts a larger spectral index $n_s$ and a smaller tensor-to-scalar ratio $r$, 
\be
V (\chi) = m^4 (1-\cos \chi/f)^\frac{\nx}{2}
\label{eqn:Vnew}
\ee
with $0<\nx<1$, we have been able to identify a family of models with viable spectra. Moreover, the model predicts a blue-tilted gravitational wave spectrum that is dominated by one handedness of circular polarization, resulting in unique observational targets for cosmic microwave background and gravitational wave experiments. The handedness of the gravitational wave spectrum is transferred to a chiral asymmetry of leptons through the axial-gravitational anomaly. Requiring that this asymmetry matches the observed baryon asymmetry of the Universe, we obtain a novel constraint on our model which places the CMB B-mode spectrum squarely within reach of ongoing and future experiments.
 
These results are timely because CMB experiments have recently ruled out the simplest inflationary scenarios that predicted high amplitude B-modes as a consequence of a high energy scale of inflation \cite{Ade:2015tva}. Our model is one of a new class of recently proposed gauge field models \cite{Dimastrogiovanni:2016fuu,Adshead:2016omu,Adshead:2017hnc,Maleknejad:2016qjz,Agrawal:2017awz} which illustrate that a detectable B-mode signal can be generated from inflation at a relatively low energy scale \cite{Fujita:2017jwq}.  
 
In the following sections we introduce the model and present our calculation procedure for the scalar and tensor spectra. Our first main result is summarized in Fig.~\ref{fig:nsr}, which shows the range of our family of models in the $n_s-r$ parameter space. We examine the unique features imprinted on the CMB temperature and polarization anisotropy spectra in these models, and look ahead to forecast the ability of future gravitational wave observatories to corroborate this model. Our second main result, the frequency spectrum of the gravitational wave background, is shown in Fig.~\ref{fig:gwspectrum}. Finally, we examine a possible connection to leptogenesis. Fig.~\ref{fig:baryo} illustrates our third main result, whereby models that explain the matter-antimatter asymmetry of the Universe also predict an approximate lower bound on the tensor-to-scalar ratio.

\section{The Theory}

The action for the theory is
\be
	S = \int d^4 x \sqrt{-g} \left( \frac{M_P^2}{2}R-\frac{1}{4}F_{a\mn}F^{a \mn} - \frac{1}{2}(\partial \chi)^2 -V(\chi) + \frac{\chi}{M}F_{a\mn} \tilde{F}^{a\mn} + {\cal L}_{m} \right),
	\label{eqn:thy}
\ee
where we use metric signature $-+++$ and curvature conventions as in Ref.~\cite{Wald:1984rg}. The SU(2) gauge field is defined by the field strength tensor 
\be
	F^{a}_\mn \equiv \partial_\mu A^{a}_\nu - \partial_\nu A^{a}_\mu -g \epsilon^{abc}A_{b\mu} A_{c\nu},
\ee
where $g$ is the coupling constant, and the dual field-strength tensor is
\be
	\tilde{F}^{a\mn} = \frac{1}{2}\epsilon^{\mu\nu\alpha\beta} F^{a}_{\alpha\beta}.
\ee
Greek letters are used to represent space-time indices, and Latin letters $i,\,j,\, ...$ are used for spatial indices. The SU(2) indices are indicated by $a,b,c,...$, and are raised and lowered by a metric $\eta = {\rm diag}(1,1,1)$. The permutation density is $\epsilon^{\mu\nu\alpha\beta} = [\mu\nu\alpha\beta]/\sqrt{-g}$ and $[0123]=+1$. We use the potential of Eq.~(\ref{eqn:Vnew}) with $0<\nx<1$, which is symmetric under shifts $\chi \to \chi + 2 \pi f$. Although this potential is too steep to yield slow roll inflation, the axionic coupling to the SU(2) field sufficiently flattens the potential. The cusp at the origin is not important for this model, and may be safely smoothed off; alternatively, the cusp might play a role in the post-inflationary reheating phase. For simplicity, however, we work with the potential
\begin{equation}
V =m^{4}(\chi/m)^\nx /\nx
\label{eqn:Vpot}
\end{equation}
which has the benefit of introducing one fewer parameter. We comment on the differences with this potential in later discussion.

The background cosmology, in a Robertson-Walker spacetime with line element $ds^2 = a(\tau)^2(-d\tau^2 + d\vec x^2)$, consists of a homogeneous scalar $\chi(\tau)$ and vector field $A^{b}_\mu$ in a flavor-space locked configuration, 
\be
	A_i^b = \phi (\tau) \delta^b_i
\ee
with all other components vanishing. The non-zero components of the field strength tensor are
\be
F^b_{0i} =\frac{\phi'}{a} \delta^b_i\, \qquad F^b_{ij} = -g \phi^2 {\epsilon^b}_{ij}
\ee
where derivatives w.r.t.~conformal time are denoted with a prime. This field configuration resembles a pair of uniform, stationary electric and magnetic fields for each flavor, pointing along the $x-,\, y-,\, z-$directions. Although the configuration is anisotropic in flavor, it is isotropic in pressure and energy. 
The energy density and pressure are
\begin{equation}
\rho = \frac{3}{2 a^4}\left(\phi'^2 + g^2 \phi^4\right) + \frac{1}{2}\left(\frac{\chi'}{a}\right)^2 + V, \qquad
p = \frac{1}{2 a^4}\left(\phi'^2 + g^2 \phi^4\right) + \frac{1}{2}\left(\frac{\chi'}{a}\right)^2 - V.
\end{equation}
The equations of motion are
\begin{equation}
\chi'' + 2 \frac{a'}{a}\chi' + a^2 V_{,\chi} = 12 \frac{g}{a^2 M} \phi^2\phi', \qquad
\phi'' + 2g^2 \phi^3 + 4 g \phi^2 \frac{\chi'}{M}=0.
\end{equation}
The free parameters of the theory are $g$, $M$, and the parameters of the potential. (We have also considered enlarging the gauge group, although no new behavior emerges. See Appendix~\ref{app:SUN}.)

The new accelerating solutions occur when the field $\chi$ sits at the extremum of the effective potential $V_{\rm eff} = V - \frac{\chi}{M}F_{a\mu\nu}\tilde F^{a\mu\nu}$. We refer to the solution, wherein $\partial V_{\rm eff}/\partial \chi=0$, as the ``accelerating track." For our numerical investigations, we may start the field evolution either on or off this track. While we have not fully investigated the phase space dynamics, we have found that a large range of initial conditions lead to inflation. The accelerating track has a finite life, however, in the sense that inflation eventually ends in these models. The track leads the scalar field to the bottom of its potential, whereupon the gauge field kinetic energy, in the form of time-evolving electric and magnetic fields, dominates with equation of state $w=1/3$. Hence, the model has a graceful exit from inflation.

\begin{figure}[t]
\includegraphics[width=0.45\linewidth]{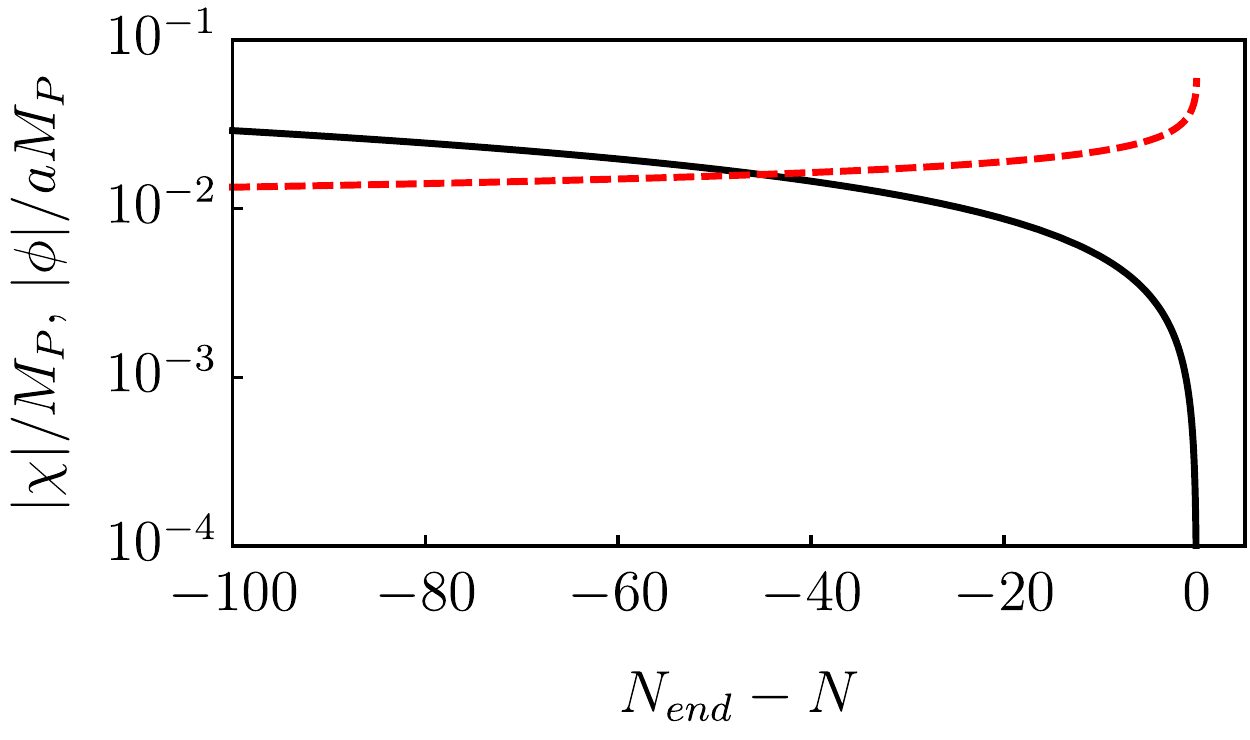}
\hspace{0.2cm}
\includegraphics[width=0.45\linewidth]{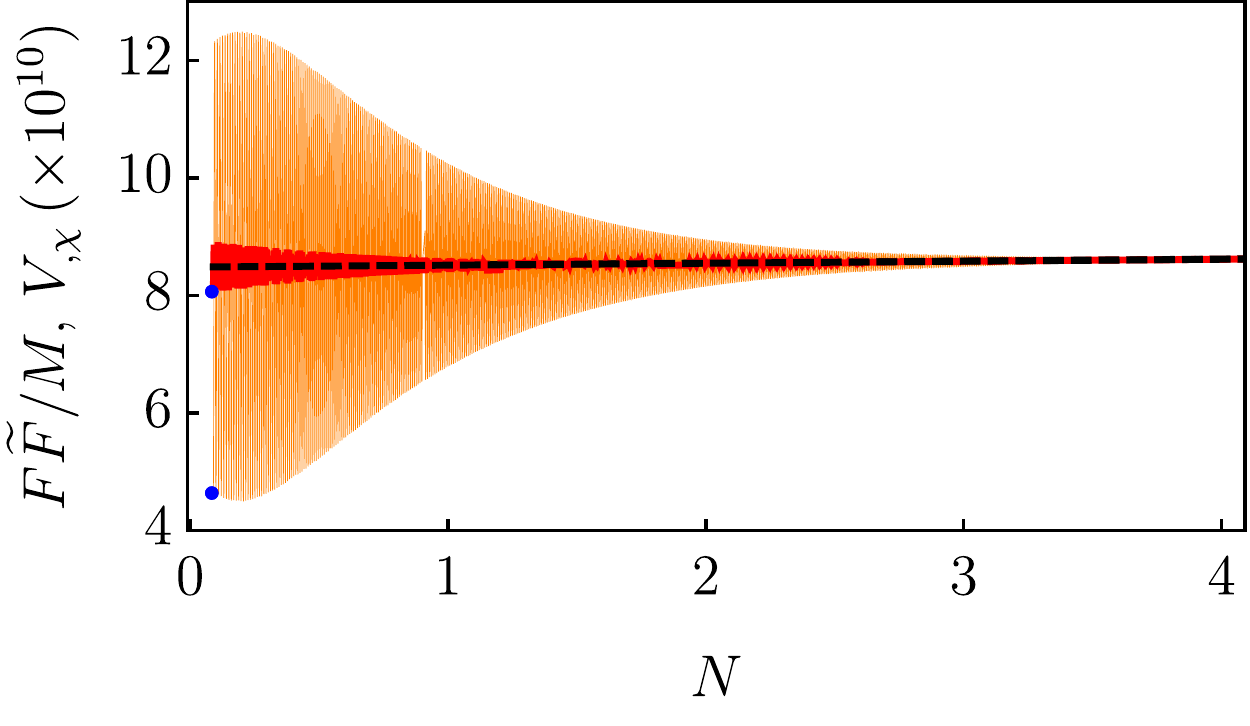}
\caption{ (Left) The evolution of the scalar field $\chi$ (solid, black) and the gauge field amplitude $\phi$ (dashed, red) are shown across 100 e-foldings of inflation, where $N=\ln a$ and $N_{\rm end}=0$ is the end of inflation. (Right) The evolution of the scalar field in terms of $V_{,\chi}$ (dashed, black) and the gauge field in terms of $F\widetilde F/M$ for two sets of initial conditions (red and orange). The horizontal axis is offset in this figure, so the end of inflation occurs at large $N$. The (blue) dot shows the starting point for the gauge field. The system rapidly evolves to the accelerating track, as shown by the convergence of the scalar and gauge field trajectories towards the right.}
\label{fig:bkgd}
\end{figure}

As an example, we show several trajectories with the same set of parameters but varying initial conditions. In Fig.~\ref{fig:bkgd} we show the evolution of $\chi$ and $\phi$ in a scenario with $\nx=1/4$. We also show the evolution of each of the two parts of the effective potential as the system evolves towards and along the accelerating track. The slow roll parameter $\epsilon_V$ based on the potential $V$ is much greater than unity, $\epsilon_V \gg 1$, so that inflation without the axionic coupling to the gauge field is not feasible. At $60$ e-foldings before the end of inflation, the standard slow roll expressions predict a curvature power spectrum amplitude $\Delta_\zeta^2 = H_*^2/(8 \pi^2\epsilon_{H*}) = 3.1 \times 10^{-10}$ with spectral index $n_s = 1+2\eta_{H*} - 4 \epsilon_{H*} = 0.98$, and a spectrum of gravitational waves with tensor-to-scalar ratio $r=16\epsilon_{H*}=0.025$. As we show in the next sections, the actual fluctuation spectra are dramatically different.

\section{Scalar Perturbations}
\label{sec:scalar}

We introduce scalar perturbations of the metric and gauge field. Our notation and procedure follows very closely that of Refs.~\cite{Dimastrogiovanni:2012ew,Namba:2013kia}. For the metric, we write
\begin{equation}
ds^2 = a^2(\tau)\left((-1+2\Phi)d\tau^2 + 2 \partial_i B d\tau \, dx^i + d\vec x^2\right).
\end{equation}
Since we will be considering only linearized perturbations there is no coupling of Fourier modes, so that we can choose the Fourier wave vector to point along the $z$-direction without any loss of generality. Hence, we consider metric perturbations with wave vector pointing in the $z$-direction, whereby the non-zero metric perturbations are $\delta g_{tt} = a^2 2 \Phi$ and $\delta g_{tz} = a^2 \partial_z B$. Likewise, for the gauge field
\begin{equation}
\delta A^a_\mu = a  \left( \begin{array}{cccc}
0 & \delta Q & 0 & 0 \cr
0 & 0 &\delta Q & 0 \cr
\partial_z Y & 0 & 0 & \delta Q + \partial_z^2 \delta M
\end{array}\right).
\label{eqn:dAtnsr}
\end{equation}
We proceed to insert these into Eq.~(\ref{eqn:thy}) and evaluate the second order action. The Fourier transformed action is
\begin{equation}
S = \int d^3k \, d\tau L, \qquad L =  
X^{\dagger\prime} A X^\prime + (X^{\dagger\prime} B X + {h.c.}) + X^{\dagger} C X +  (X^{\dagger\prime} D N + {h.c.}) + (X^{\dagger} E N + {h.c.}) + N^\dagger F N
\label{eqn:scalaraction}
\end{equation}
where $X = \{\delta M,\, \delta Q,\, \delta\chi \}$ are the dynamical degrees of freedom, and $N=\{Y,\, B,\, \Phi\}$ are the constraints. 
After evaluating the constraint equations and re-inserting into the action, we have
\begin{equation}
L = X^{\dagger\prime} (A- D F^{-1} D^\dagger) X^\prime + X^{\dagger\prime}(B-D F^{-1} E^\dagger) X 
+ X^{\dagger}(B^\dagger-E F^{-1} D^\dagger) X^\prime + X^{\dagger} (C- E F^{-1} E^\dagger) X.
\end{equation}
Now that the constraint or gauge variables have been eliminated, we can obtain the equations of motion for the scalar perturbations.

To obtain a canonical kinetic term, we introduce the transformation matrix ${\cal M}$ so that $X = {\cal M}\Delta$, and $\Delta = \{\Delta_1,\,\Delta_2,\,\Delta_3\}$ are the new scalar modes. The Lagrangian becomes
\begin{eqnarray}
L &=& \frac{1}{2} \left( \Delta^{\dagger\prime} T \Delta^\prime + \Delta^{\dagger\prime} K_1 \Delta + \Delta^{\dagger} K_2 \Delta^\prime + \Delta^\dagger W \Delta \right) \\
\frac{1}{2}T &=&  {\cal M}^\dagger (A- D F^{-1} D^\dagger){\cal M} \\
\frac{1}{2}K_1 &=& {\cal M}^\dagger (A- D F^{-1} D^\dagger) {\cal M}^\prime + {\cal M}^\dagger(B- D F^{-1} E^\dagger) {\cal M}  \\
\frac{1}{2}K_2 &=& {\cal M}^{\dagger\prime} (A- D F^{-1} D^\dagger) {\cal M} + {\cal M}^\dagger(B^\dagger- E F^{-1} D^\dagger) {\cal M}  \\
\frac{1}{2}W &=& {\cal M}^{\dagger\prime} (A- D F^{-1} D^\dagger) {\cal M}^\prime + {\cal M}^\dagger (C- E F^{-1} E^\dagger) {\cal M} \cr
&& + {\cal M}^{\dagger\prime}(B-D F^{-1} E^\dagger) {\cal M} + {\cal M}^{\dagger}(B^\dagger - E F^{-1} D^\dagger) {\cal M}^\prime.
\end{eqnarray} 
We can integrate by parts so that the Lagrangian simplifies to
\begin{eqnarray}
L &=& \frac{1}{2} \left( \Delta^{\dagger\prime} T \Delta^\prime + \Delta^{\dagger\prime} K \Delta - \Delta^{\dagger} K \Delta^\prime - \Delta^\dagger \Omega^2 \Delta \right) 
\label{eqn:LDelta}\\
K &=& \frac{1}{2}(K_1 - K_2) \\
\Omega^2 &=& -W + \frac{1}{2}(K_1 + K_2)^\prime.
\end{eqnarray} 
The resulting equation of motion is
\begin{equation}
\Delta'' + T^{-1}(T' + 2 K) \Delta' + T^{-1}(K' + \Omega^2)\Delta = 0.
\label{eqn:DeltaEOM}
\end{equation}
By suitable choice of ${\cal M}$ we can arrange that in the high frequency limit,
\begin{equation}
T = \mathbb{I}, \qquad K = {\cal O}(k^0), \qquad \Omega^2 = k^2 \mathbb{I},
\end{equation}
so that each mode $\Delta$ behaves like a free oscillator.

\subsection{Quantum Fluctuations}

The action of the normal modes $\Delta$ of the scalar perturbations of the metric and gauge field at high frequency resembles that of free fields in Minkowski spacetime. As in the standard treatment, we promote these modes and conjugate momentum to quantum operators:
\begin{eqnarray}
\Delta^\ell \to \hat \Delta^\ell &=& \int \frac{d^3k}{(2 \pi)^3}\left[ \Delta^\ell_k(\tau) \hat a_{\vec k}^\ell e^{i\vec k \cdot \vec x} 
+ \Delta_k^{\ell*}(\tau) \hat a_{\vec k}^{\ell\dagger}  e^{-i\vec k \cdot \vec x}\right] \\
\Delta^{\ell\prime} \to \hat \pi^\ell_{\Delta} &=& \int \frac{d^3k}{(2 \pi)^3}\left[ \Delta_k^{\ell\prime}(\tau) \hat a_{\vec k}^\ell e^{i\vec k \cdot \vec x} + \Delta_k^{\ell\prime *}(\tau) \hat a_{\vec k}^{\ell\dagger} e^{-i\vec k \cdot \vec x}\right]
\end{eqnarray}
where the superscript $\ell=1,\,2,\,3$ distinguishes among the three normal modes. We can apply the canonical commutation relations on the field and its conjugate momentum, $[\hat \Delta^\ell(\vec x),\,\hat\pi^\ell_{\Delta}(\vec x')] = i \delta(\vec x - \vec x')$ and similarly enforce the normalization of the annihilation and creation operators $[\hat a_{\vec k}^\ell,\, \hat a_{\vec k'}^{\ell\dagger}]= (2\pi)^3 \delta(\vec k - \vec k')$, whereupon the mode functions are normalized by the condition $i(\Delta_k^{\ell*} \Delta_k^{\ell\prime} - \Delta_k^{\ell*\prime}\Delta_k^\ell) = 1$. Since the solution to the mode function wave equation at high frequency is $\Delta^\ell_k \propto e^{-i k \tau}$, we can start modes that are sufficiently far inside the horizon that $k \gg a'/a,\, g\phi,\, \chi'/M$ with the initial condition $\Delta^\ell_k |_i =  e^{-i k \tau}/\sqrt{2 k}$, $\Delta_k^{\ell\prime} |_i =-i k e^{-i k \tau}/\sqrt{2 k}$.  The annihilation and creation operators for different normal modes, e.g. with different values of $\ell$, all commute. So, our three quantum fields live on a product of three separate Hilbert spaces, and we will need to evolve the coupled equations of motion Eq.~(\ref{eqn:DeltaEOM}) under three separate sets of initial conditions. The first we will call the ${\mathscr H}_1$ system, under which we evolve Eq.~(\ref{eqn:DeltaEOM}) with initial conditions
\begin{equation}
{\mathscr H}_1: \Delta^1 |_i = \frac{e^{-i k \tau_i}}{\sqrt{2 k}}, \quad \Delta^{1\prime} |_i =-i k \frac{e^{-i k \tau_i}}{\sqrt{2 k}}, \quad \Delta^2 |_i = \Delta^{2\prime} |_i  = 0, \quad \Delta^3 |_i = \Delta^{3\prime} |_i  = 0.
\label{eqn:initDelta1}
\end{equation}
The second and third follow similarily:
\begin{equation}
{\mathscr H}_2: \Delta^2 |_i = \frac{e^{-i k \tau_i}}{\sqrt{2 k}}, \quad \Delta^{2\prime} |_i =-i k \frac{e^{-i k \tau_i}}{\sqrt{2 k}}, \quad \Delta^1 |_i = \Delta^{1\prime} |_i  = 0, \quad \Delta^3 |_i = \Delta^{3\prime} |_i  = 0.
\label{eqn:initDelta2}
\end{equation}
\begin{equation}
{\mathscr H}_3: \Delta^3 |_i = \frac{e^{-i k \tau_i}}{\sqrt{2 k}}, \quad \Delta^{3\prime} |_i =-i k \frac{e^{-i k \tau_i}}{\sqrt{2 k}}, \quad \Delta^1 |_i = \Delta^{1\prime} |_i  = 0, \quad \Delta^2 |_i = \Delta^{2\prime} |_i  = 0.
\label{eqn:initDelta3}
\end{equation}
The power spectrum of fluctuations in each of the normal modes is then the sum over the three Hilbert spaces,
$|\Delta^\ell|^2 = |\Delta^\ell|^2_{{\cal H}_1} + |\Delta^\ell|^2_{{\cal H}_2} + |\Delta^\ell|^2_{{\cal H}_3}$ for each $\ell$.

\subsection{Curvature Spectrum}

The ultimate goal in evaluating the scalar perturbations is to determine the power spectrum of curvature fluctuations. From the
definition of the curvature fluctuation
\begin{equation}
\zeta = - \frac{H}{\dot\rho}\delta\rho,
\end{equation}
where $\delta\rho = -\delta T^t_t$, we arrive at the following result:
\begin{equation}
\zeta = \left(R_1 X' + R_2 X + R_3 N\right). \\
\end{equation}
Using results from the previous subsections, we determine that
\begin{equation}
\zeta = (R_1 - R_3 F^{-1}D){\cal M}\Delta' + \left[ (R_1 - R_3 F^{-1} D){\cal M}' + (R_2 - R_3 F^{-1} E){\cal M}\right]\Delta.
\end{equation}
This must be evaluated at the end of inflation, for the  curvature perturbation on each Hilbert space. The power spectrum is
\begin{equation}
\Delta_\zeta^2 = \frac{k^3}{2 \pi^2}\left( |\zeta|^2_{{\mathscr H}_1} +  |\zeta|^2_{{\mathscr H}_2}+  |\zeta|^2_{{\mathscr H}_3} \right).
\label{eq:deltazeta}
\end{equation}
For comparison, in the standard case of single field slow roll inflation, the power spectrum is $\Delta_\zeta^2 =H_*^2/8 \pi^2 \epsilon_{H*}$ with spectral index $n_s = 1 + d \ln\Delta_\zeta^2/d \ln k = 1 + 2 \eta_{H*} - 4 \epsilon_{H*}$.

\section{Tensor Modes}
\label{sec:TensorModes}

We consider gravitational waves and tensor fluctuations of the gauge field. Since we will be considering only linearized perturbations, there is no coupling of Fourier modes, so that we can choose the Fourier wave vector to point along the $z$-direction without any loss of generality. Hence, we consider a gravitational wave propagating in the $z$-direction:
\begin{equation}
\delta g_{\mu\nu} = a^2 h_{\mu\nu} = a^2 \left(\begin{array}{cccc}
0 & 0 & 0 & 0 \cr
0 & h_+ & h_\times & 0 \cr
0 & h_\times & -h_+ & 0 \cr
0 & 0 & 0 & 0 
\end{array}\right).
\end{equation}
Similarly, we consider a $z$-directed gauge field wave
\begin{equation}
\delta A^a_\mu = a\, t^a_\mu = a \left( \begin{array}{cccc}
0 & t_+ & t_\times & 0 \cr
0 & t_\times & -t_+ & 0 \cr
0 & 0 & 0 & 0
\end{array}\right).
\label{eqn:Atnsr}
\end{equation}
These expressions are inserted into the action (\ref{eqn:thy}), which is then expanded to quadratic order. In order that the gravitational wave and gauge field have canonical kinetic terms, we introduce a change of variables
\begin{equation}
h_{+,\,\times} = \frac{\sqrt{2}}{a M_P}v_{+,\,\times}, \qquad t_{+,\,\times} = \frac{1}{\sqrt{2} a}u_{+,\,\times}.
\end{equation} 
However, the $+$ and $\times$ polarization modes are coupled.
Hence, by rotating into a circular polarization basis
\begin{eqnarray}
&& v_{+} = \frac{1}{\sqrt{2}}(v_L + v_R),\quad v_{\times} = \frac{i}{\sqrt{2}}(v_L - v_R), \cr
&& u_{+} = \frac{1}{\sqrt{2}}(u_L + u_R),\quad u_{\times} = \frac{i}{\sqrt{2}}(u_L - u_R),
\end{eqnarray}
the left- and right-circularly polarized equations decouple. The full equations, in terms of the Fourier space amplitudes, are as follows:
\begin{eqnarray}
&&v_L'' + \left[k^2 - \frac{a''}{a}  + \frac{2}{a^2 M_P^2}(g^2 \phi^4 - \phi'^2)\right] v_L = \frac{2}{a M_P}\left[ (g \phi + k) g  \phi^2 u_L -  \phi' u_L'\right]
\label{eqn:vL} \\
&&u_L'' + \left[k^2 + 2 g k \phi +4(g \phi + k)\frac{\chi'}{M}\right]u_L  
=\frac{2}{a M_P}\left[ a\left(\frac{v_L}{a}\right)' \phi' + g \phi^2 \left(k-g\phi +4 \frac{\chi'}{M} \right)  v_L  \right].
\label{eqn:uL}
\end{eqnarray}
The equations for $v_R,\,u_R$ are obtained by replacing $k \to -k$.

The coupled gravitational wave -- gauge field system in Eqs.~(\ref{eqn:vL}-\ref{eqn:uL}) has three notable features. First, the gravitational wave acquires an additional mass-like term 
\be
\frac{2}{a^2 M_P^2}(g^2 \phi^4 - \phi'^2)
\ee
arising from the anisotropic shear of the gauge field. Second, a tachyonic instability in the left-circularly polarized gauge field wave, occurring when 
\be
k^2 + 2 g k \phi +4(g \phi + k)\frac{\chi'}{M} < 0
\ee
breaks chiral symmetry and pumps energy into left-circularly polarized gravitational waves for $g>0$, $\phi<0$, and $\chi'<0$. In contrast, for the same parameter signs, the right-circular polarization has no such instability. Third, at high frequencies $k \gg a'/a,\, g\phi,\, \chi'/M$, the gravitational wave and gauge field interconvert through the phenomenon of gravitational wave -- gauge field oscillations \cite{Caldwell:2016sut}. All three effects play a role in the production of a primordial gravitational wave spectrum.

\subsection{Quantum Fluctuations}

The action of the gravitational wave and gauge field at high frequency resembles that of a free field in Minkowski spacetime. As in the standard treatment, we promote the gravitational wave and its conjugate momentum to quantum operators:
\begin{eqnarray}
v \to \hat v &=& \int \frac{d^3k}{(2 \pi)^3}\left[ v_k(\tau) \hat a_{\vec k} e^{i\vec k \cdot \vec x} 
+ v_k^{*}(\tau) \hat a_{\vec k}^\dagger  e^{-i\vec k \cdot \vec x}\right] \\
v' \to \hat \pi_v &=& \int \frac{d^3k}{(2 \pi)^3}\left[ v'_k(\tau) \hat a_{\vec k} e^{i\vec k \cdot \vec x} + v_k^{\prime *}(\tau) \hat a_{\vec k}^\dagger e^{-i\vec k \cdot \vec x}\right].
\end{eqnarray}
In this case, we can apply the canonical commutation relations on the field and its conjugate momentum, $[\hat v(\vec x),\,\hat\pi_v(\vec x')] = i \delta(\vec x - \vec x')$ and similarly enforce the normalization of the annihilation and creation operators $[\hat a_v(\vec k),\, \hat a^\dagger_v(\vec k')]= (2\pi)^3 \delta(\vec k - \vec k')$, whereupon the mode functions are normalized by the condition $i(v_k^* v_k' - v_k^{*\prime}v_k) = 1$. Since the solution to the mode function wave equation at high frequency is $v_k \propto e^{-i k \tau}$, we can start modes that are sufficiently far inside the horizon that $k \gg a'/a,\, g\phi,\, \chi'/M$ with the initial condition $v_k |_i =  e^{-i k \tau}/\sqrt{2 k}$, $v'_k |_i =-i k e^{-i k \tau}/\sqrt{2 k}$ for each polarization. Next, the gauge field fluctuations are a separate, independent quantum field. Similarly, we promote the gauge field tensor wave and its conjugate momentum to quantum operators:
\begin{eqnarray}
u \to \hat u &=& \int \frac{d^3k}{(2 \pi)^3}\left[ u_k(\tau) \hat b_{\vec k} e^{i\vec k \cdot \vec x} + u_k^{*}(\tau) \hat b_{\vec k}^\dagger  e^{-i\vec k \cdot \vec x}\right] \\
u' \to \hat \pi_u &=& \int \frac{d^3k}{(2 \pi)^3}\left[ u'_k(\tau) \hat b_{\vec k} e^{i\vec k \cdot \vec x} + u_k^{\prime *}(\tau) \hat b_{\vec k}^\dagger e^{-i\vec k \cdot \vec x}\right].
\end{eqnarray}
As above, we find that for modes that are sufficiently far inside the horizon, the appropriate initial conditions are $u_k |_i =  {e^{-i k \tau}}/{\sqrt{2 k}}$, $u'_k |_i =-i k  {e^{-i k \tau}}/{\sqrt{2 k}}$, for each polarization. But the gauge field tensor modes are built on an independent Hilbert space from the gravitational waves. Hence, the $\hat a$ and $\hat b$ annihilation and creation operators commute with each other. So in fact we need two copies of Eqs.~(\ref{eqn:vL}-\ref{eqn:uL}). 

The first copy we will call the ${\mathscr H}_v$ system, to describe the evolution of gravitational and gauge field tensor waves due to quantum fluctuations of the {\it gravitational} wave vacuum. The initial conditions of this system are summarized as follows:
\begin{equation}
{\mathscr H}_v: v_L |_i = \frac{e^{-i k \tau_i}}{\sqrt{2 k}}, \quad v'_L |_i =-i k \frac{e^{-i k \tau_i}}{\sqrt{2 k}}, \quad u_L |_i = u'_L |_i = 0.
\label{eqn:initv}
\end{equation}
The second copy is the ${\mathscr H}_u$ system, which describes the evolution of gravitational and gauge field tensor waves due to quantum fluctuations of the {\it gauge field tensor} wave vacuum:
\begin{equation}
{\mathscr H}_u:  v_L |_i = v'_L |_i = 0,  \quad u_L |_i = \frac{e^{-i k \tau_i}}{\sqrt{2 k}}, \quad u'_L |_i =-i k \frac{e^{-i k \tau_i}}{\sqrt{2 k}}.
\label{eqn:initu}
\end{equation}
The right-circular polarization modes follow a similar procedure with identical initial conditions; only the equation of motion differs. The power spectrum of fluctuations is
\begin{eqnarray}
\langle \hat h_{ij\,L,\vec k}(\tau)   \hat h^{ij}_{L,\vec k'}(\tau)  \rangle &=&
2\langle \hat h_{L,\vec k}(\tau)   \hat h_{L,\vec k'}(\tau)  \rangle =
2\left(\langle {\mathscr H}_v | \hat h_{L,\vec k}(\tau)   \hat h_{L,\vec k'}(\tau) | {\mathscr H}_v \rangle
+\langle {\mathscr H}_u | \hat h_{L,\vec k}(\tau)   \hat h_{L,\vec k'}(\tau) | {\mathscr H}_u \rangle  \right)  \\
&=& (2 \pi)^3 \delta(\vec k + \vec k') P_{L}(k)\\
P_{L}(k) &=& 2 \left(  | h_{L,k}|^2_{{\mathscr H}_v} + | h_{L,k}|^2_{{\mathscr H}_u}  \right)
=\frac{4}{a^2 M_P^2} \left(  | v_{L,k}|^2_{{\mathscr H}_v} + | v_{L,k}|^2_{{\mathscr H}_u}  \right)\\
\Delta^2_L(k)  &=& \frac{k^3}{2 \pi^2} P_L.
\end{eqnarray}
This procedure allows us to calculate the gravitational wave amplitude arising from quantum fluctuations of the gravitational field, and from quantum fluctuations of the gauge field, separately. There is no interference cross term between the two because they are independent quantum fields. The factor of two appearing in the power spectrum is due to our convention for the polarization tensor.

The phenomenon of gravitational wave -- gauge field oscillations \cite{Caldwell:2016sut} can be seen in the behavior of $| h_{L,k}|^2_{{\mathscr H}_v}$ and $ | h_{L,k}|^2_{{\mathscr H}_u}$. During inflation, modes deep inside the horizon oscillate with frequency $k$. However, the amplitude $| h_{L,k}|^2_{{\mathscr H}_v}$ also modulates. On its own this would be troubling, since the quantum state would appear to remember its initial conditions through its modulation phase. However, the sum $| h_{L,k}|^2+| t_{L,k}|^2_{{\mathscr H}_v}$ is constant. The same holds true for the fields on the ${\mathscr H}_u$ Hilbert space. Because deep inside the horizon, $t_{L,k}$ on ${\mathscr H}_v$ and $h_{L,k}$ on ${\mathscr H}_u$ are the same inhomogeneous mode, and likewise for $t_{L,k}$ on ${\mathscr H}_u$ and $h_{L,k}$ on ${\mathscr H}_v$, then the sum $| h_{L,k}|^2_{{\mathscr H}_v} + | h_{L,k}|^2_{{\mathscr H}_u}$ is also constant, deep inside the horizon. This erases the modulation phase, and thereby preserves the desirable feature of the inflationary quantum state.
 
Note that in the standard case we have only quantum fluctuations of the gravitational field, so that $v_L$ satisfies the equation 
\begin{equation}
v_L'' + (k^2 - a''/a)v_L=0
\end{equation} 
with initial conditions $v_L|_i = e^{-i k \tau_i}/\sqrt{2 k}$,  $v'_L|_i = -i k v_L|_i$, and the power spectrum is 
\begin{equation}
\Delta^2_L(k)  = \frac{k^3}{2 \pi^2} P_L = \frac{k^3}{2 \pi^2}   \frac{4}{a^2 M_P^2}| v_{L,k}|^2 = \frac{H_*^2}{\pi^2} .
\end{equation} 
where $H_*$ is the Hubble factor at horizon exit of a mode with comoving wavenumber $k = a_* H_*$.
An identical prescription holds for the right-circular polarization waves. 

The total tensor power spectrum is
\begin{equation}
\Delta_{GW}^2(k) =  \Delta_{L}^2(k)+\Delta_{R}^2(k). 
\end{equation}
Because the parity between left- and right-circular polarizations is broken, we introduce the V Stokes parameter or circular polarization power spectrum,  defined as
\begin{equation}
\Delta_{GW,V}^2(k) =  \Delta_{L}^2(k)-\Delta_{R}^2(k). 
\end{equation}
It is also useful to characterize the degree of parity violation in terms of a chirality parameter, $\Delta\chi = \Delta_{GW,V}^2/\Delta_{GW}^2$.

\section{Perturbation Spectra}

\begin{figure}[h]
\includegraphics[width=0.45\linewidth]{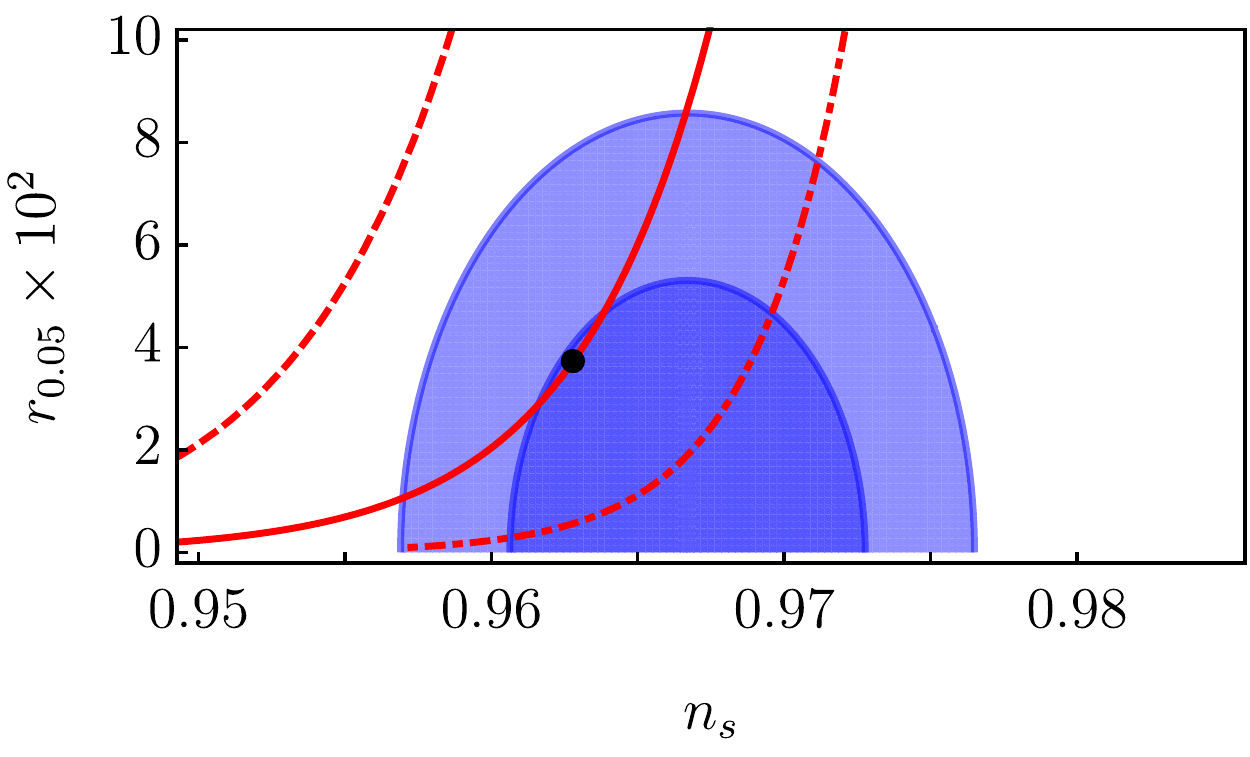}
\includegraphics[width=0.45\linewidth]{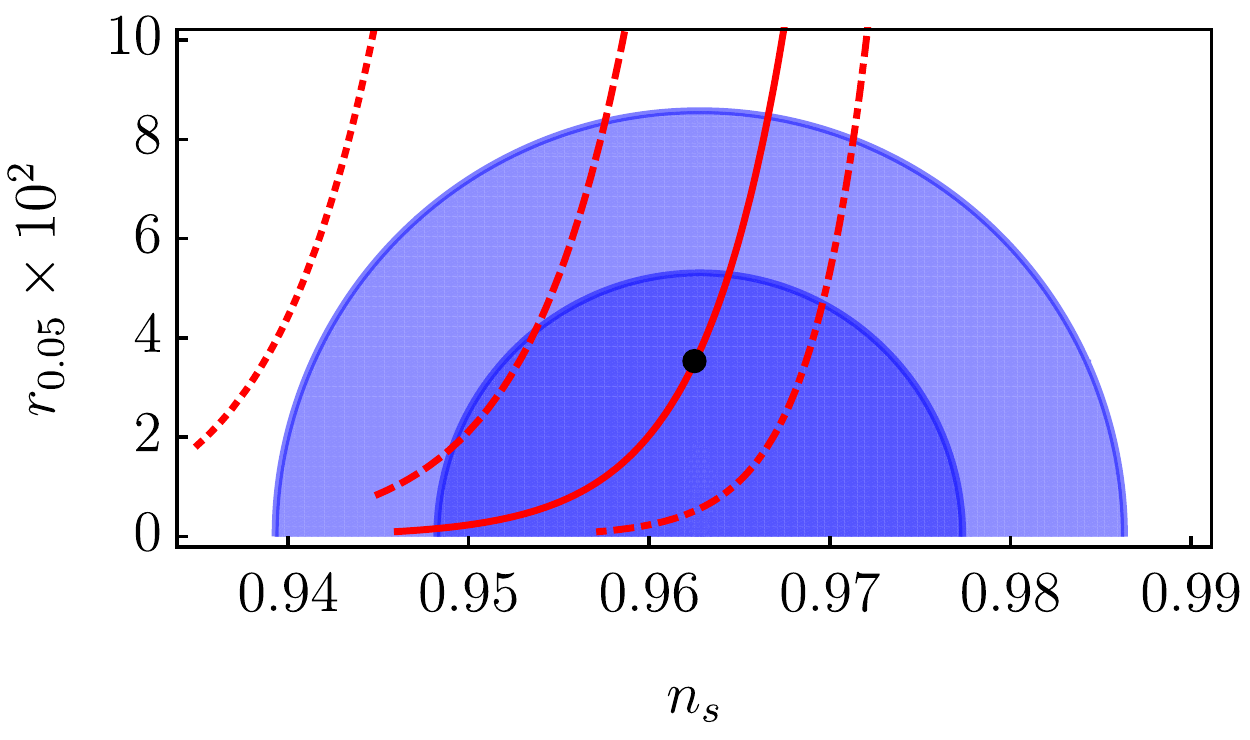}
\caption{(Left) The $\nx=1/4$ (dashed), $\nx=1/8$ (solid), $\nx = 1/16$ (dot-dashed) family of models in the $n_s - r_{0.05}$ parameter space is shown. The black dot represents the example model described in the text. The blue contours give an approximate representation of the $1,\,2\sigma$ contours based on the limits $n_s = 0.9667 \pm 0.0040\, (1\sigma)$ \cite{Ade:2015xua} and $r_{0.05} < 0.07 \,(95\%\,C.L.)$ \cite{Array:2015xqh}. (Right) We show the same curves, plus the case $\nx=1/2$ (dotted), using the relaxed constraint $n_s=0.9628\pm0.0096\, (1\sigma)$ \cite{DiValentino:2016ucb}, obtained by allowing the presence of a dark radiation component.}
\label{fig:nsr}
\end{figure}

We have evaluated the scalar and tensor spectra for a variety of axionic gauge field inflation scenarios with potential given by Eq.~(\ref{eqn:Vpot}). For each family of potentials with a given $\nx$, we choose parameters $g$, $M$, and $m$ to produce a scalar spectrum with amplitude $\Delta_\zeta^2 = 2.2\times 10^{-9}$ at a reference wavenumber $k=0.05$~inv-Mpc which exited the horizon $N_*$ e-foldings before the end of inflation. This normalization is in rough agreement with current bounds on the amplitude of the fluctuation power spectrum \cite{Ade:2015lrj}. The value of $N_*$ is determined according to a standard calculation in which adiabatic evolution of the relativistic degrees of freedom of the cosmological fluid is assumed from the end of inflation to the present day. Because our models generally predict a lower value of $H_*$ than in standard slow roll models, the e-folding calculation typically gives $N_* \simeq 55$. The relationship between the power spectrum amplitude $\Delta_\zeta^2$ and the Hubble parameter $H_*$ is nonlinear, so the range of e-foldings is more tightly constrained than in standard slow-roll inflationary models. The remaining parameter freedom yields a trajectory in the $n_s - r$ plane. We plot several such trajectories against the current bounds in Fig.~\ref{fig:nsr}. The broad overlap indicates that these models produce viable scalar and tensor spectra.

\begin{figure}[h]
\includegraphics[width=0.45\linewidth]{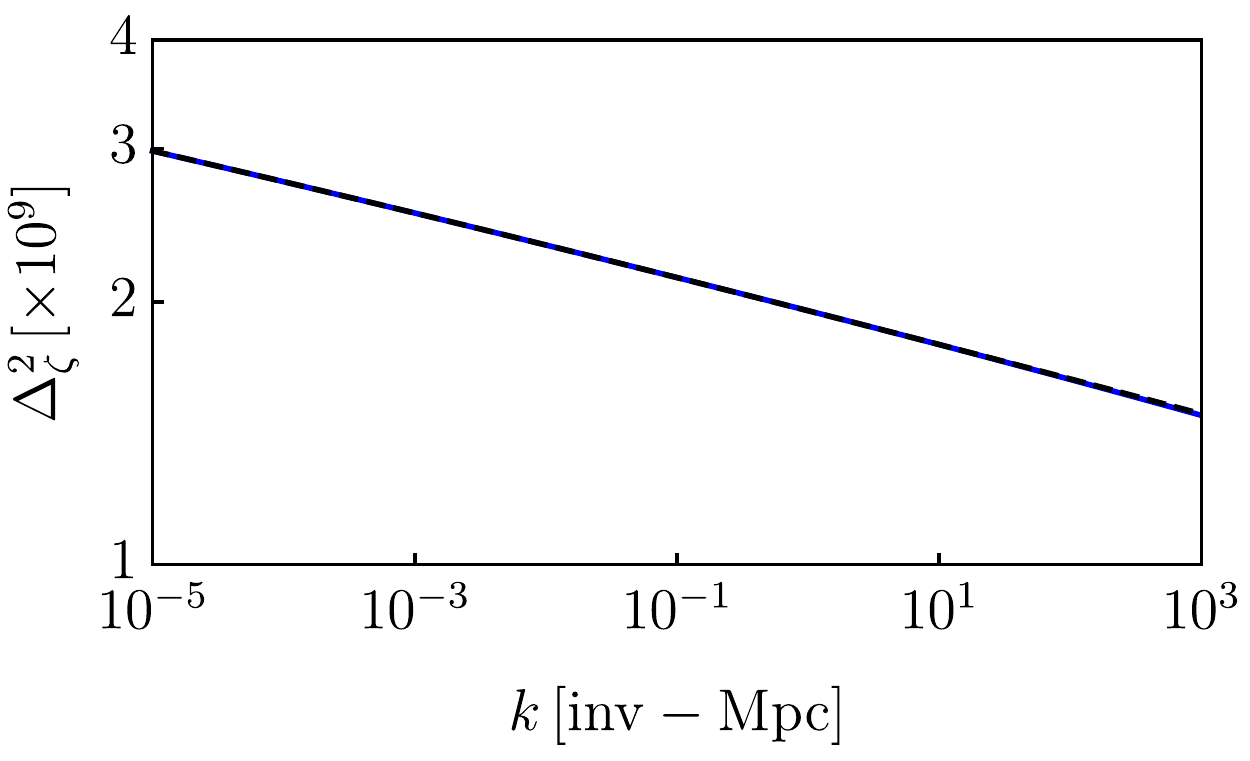}
\includegraphics[width=0.49\linewidth]{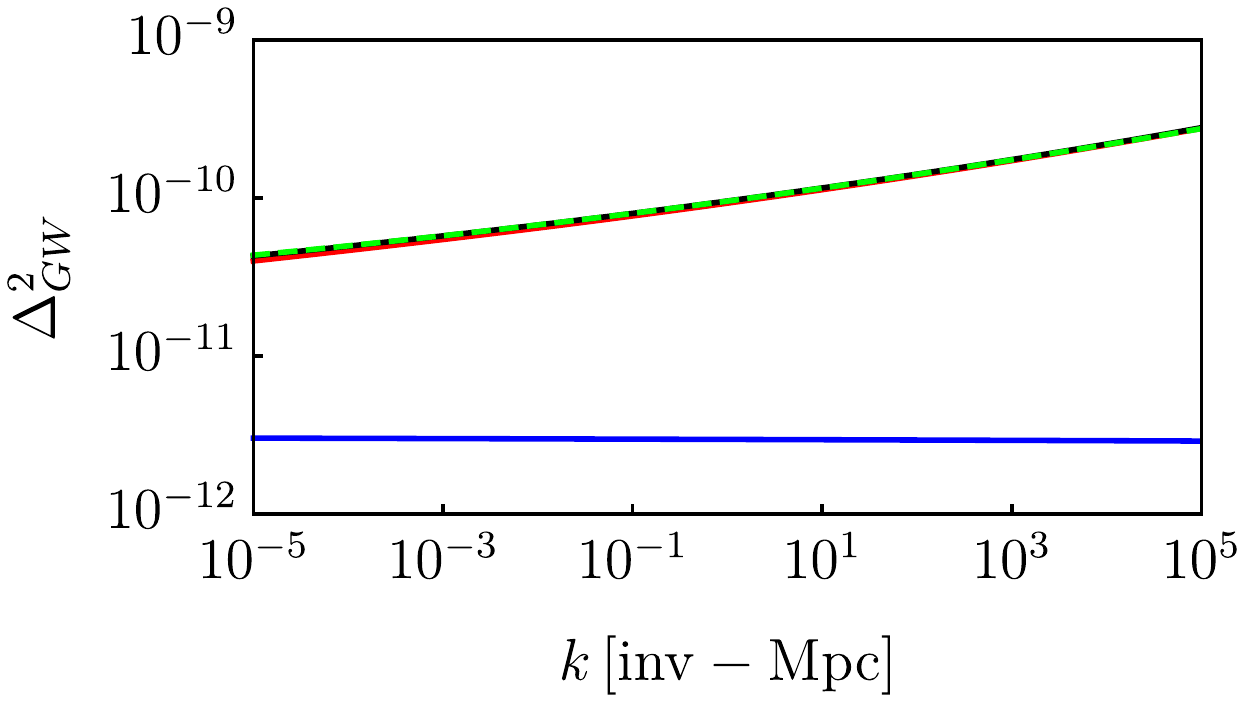}
\caption{(Left) The scalar curvature power spectrum, $\Delta_\zeta^2$, is shown as a function of wavenumber. The solid (blue) curve shows the result of our numerical calculation; the dashed (black) curve is an analytic fit with tilt $n_s =0.962$ and running index $d n_s/d\ln k = - 3 \times 10^{-4}$ at a reference wavenumber $k=0.05$~inv-Mpc. (Right) The tensor spectrum is shown as a function of wavenumber. The lower (solid, blue) curve shows the right circular polarization, whereas the upper curves show the left-circular polarization, the total spectrum, and an analytic fit with tilt $n_t = 0.074$ and running $d n_t/d\ln k = 2 \times 10^{-3}$.}
\label{fig:Delta}
\end{figure}

We consider a scenario with $\nx=1/8$, $g= 1.4 \times 10^{-3}$, $M=1.7 \times 10^{-4} M_P$, and $m = 1.7 \times 10^{-3}M_P$ as a definite example, represented by the black dot in Fig.~\ref{fig:nsr}. The scalar index is $n_s=0.962$ at $55$ e-foldings before the end of inflation, which we map to the reference wavenumber $k=0.05$~inv-Mpc. The tensor spectrum has amplitude $\Delta_{GW}^2 = 7.6\times 10^{-11}$ for a tensor-to-scalar ratio $r_{0.05}=0.035$, and $r_{0.002}=0.025$. The spectrum is almost entirely composed of left-circularly polarized gravitational waves: the chirality parameter ranges from $\Delta\chi \simeq 0.9$ at $k = 0.002$~inv-Mpc, up to unity at high frequencies. These spectra, illustrated in Fig.~\ref{fig:Delta}, are consistent with current observational bounds \cite{Ade:2015lrj,Ade:2015tva,Ade:2015xua,Array:2015xqh}. We caution that some constraint analyses use $r_{0.05}$ and then extrapolate to $r_{0.002}$ using the standard, inflationary consistency relations. Since these relationships do not apply to our model, such bounds on $r_{0.002}$ are not rigorous.

Tension in the cosmological parameter constraints \cite{Riess:2016jrr} have compelled researchers to consider a wider range of cosmological models than straight up, vanilla $\Lambda$CDM. Recently, the effect of a dark radiation component on the inflationary parameters $r$ and $n_s$ was examined, in light of current data (e.g. Refs.~\cite{Calabrese:2011hg,Bernal:2016gxb,DiValentino:2016hlg,DiValentino:2016ucb}). Even though the data tightly constrain the effective number of neutrino species, $N_{\rm eff}=3.00 \pm 0.20\, (1\sigma)$, the bound on the spectral index loosens to $n_s=0.9628\pm0.0096\, (1\sigma)$ \cite{DiValentino:2016ucb}, as illustrated in the right panel of Fig.~\ref{fig:nsr}.  

We originally invoked a periodic, sinusoidal potential, but then quickly restricted our attention to a simple power law. We expect that in the case of the sinusoidal potential, the behavior of the perturbation spectra in different epochs will resemble those of the various power laws. That is, for a potential with $\nx = 1/16$ in Eq.~(\ref{eqn:Vnew}), we expect that the stages of inflation will resemble a power law as in Eq.~(\ref{eqn:Vpot}) with $\nx = 1/4$, followed by an era described by $\nx = 1/8$, and finally $\nx=1/16$. The rate at which the effective power law index varies will depend on the relative size of the inflaton $\chi$ and the mass scale $f$. We will leave the investigation of this behavior for future work.

\subsection{Cosmic Microwave Background}

\begin{figure}[h]
\includegraphics[width=0.45\linewidth]{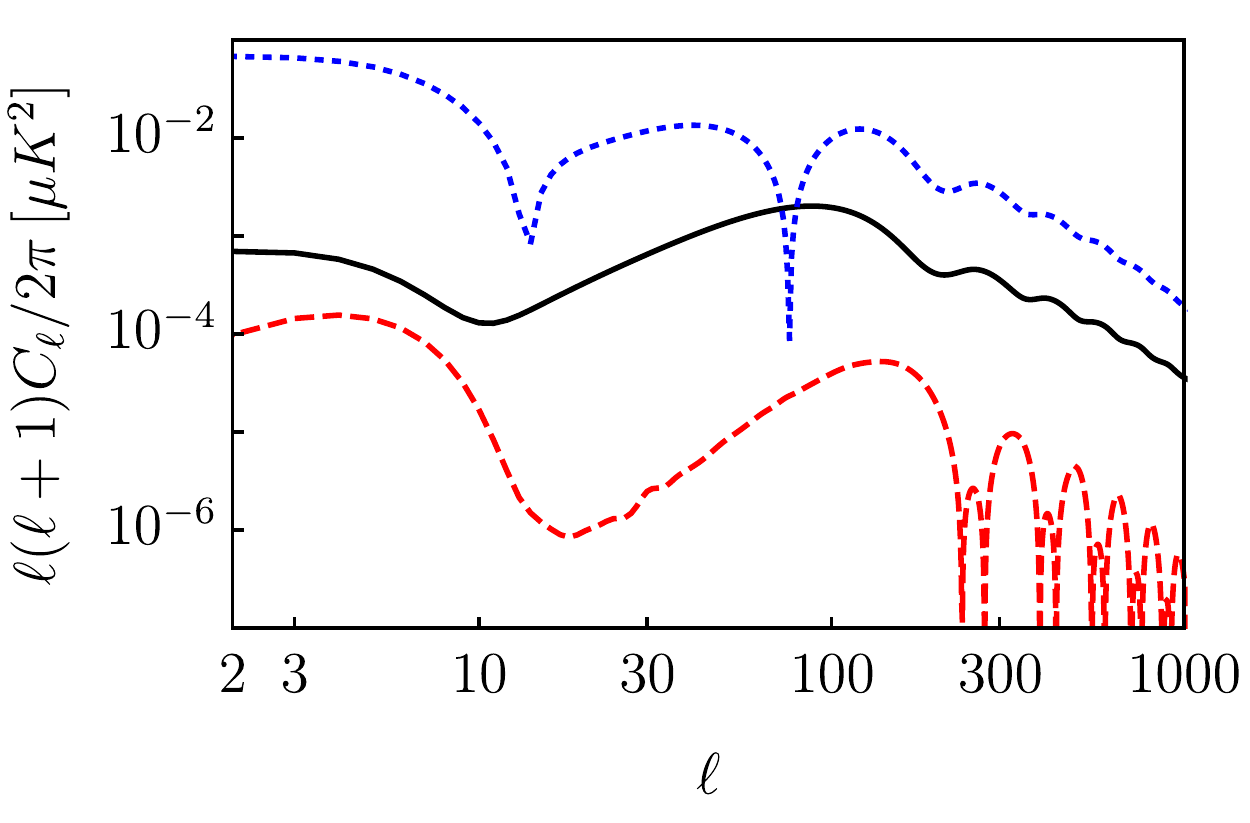}
\includegraphics[width=0.42\linewidth]{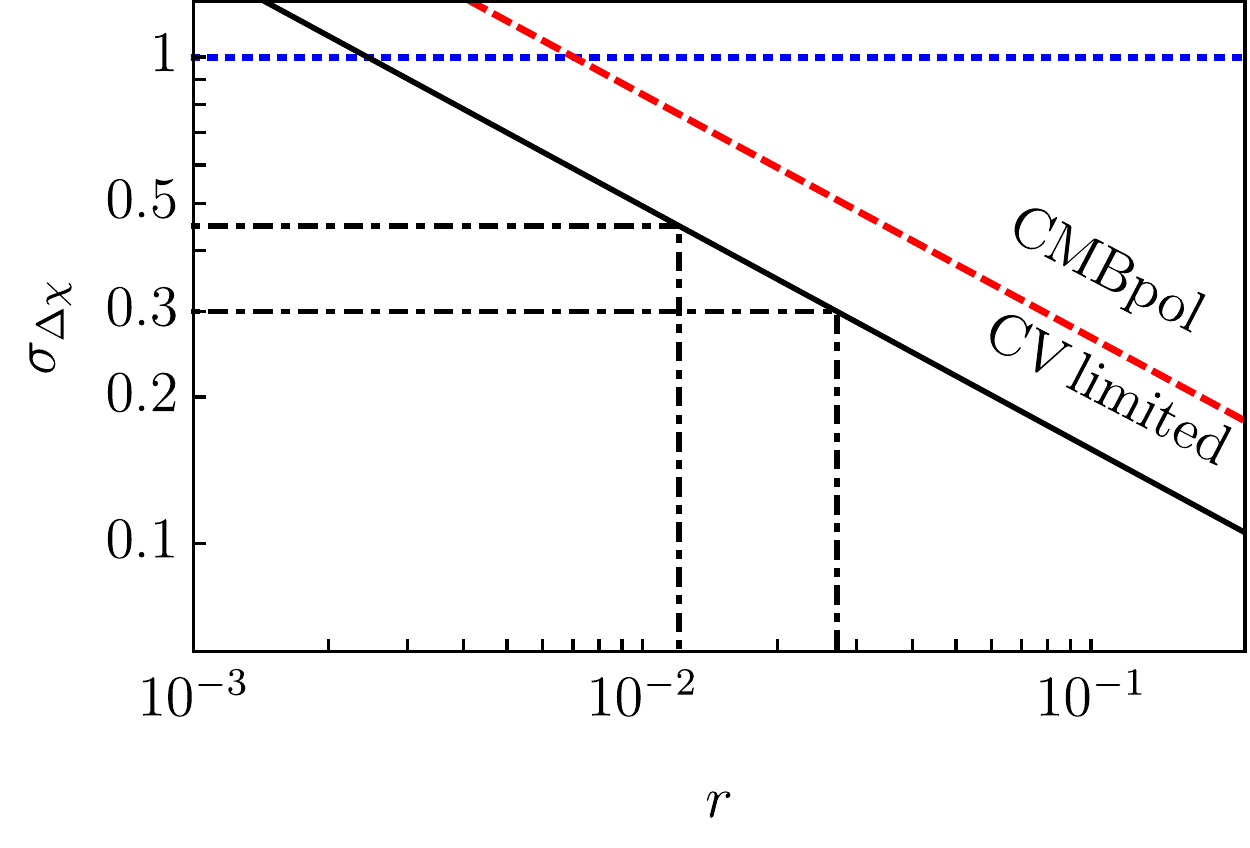}
\caption{(Left) The BB (solid, black), TB (dotted, blue) and EB (dashed, red) anisotropy power spectra are shown for our example scenario, for which $r=0.035$, $n_t = 0.074$, and $\Delta\chi = 0.92$. (Right) The sensitivity to a chiral asymmetry $\Delta\chi$ is shown as a function of the tensor-to-scalar ratio $r$ for an idealized satellite experiment (CMBpol) and a cosmic-variance limited experiment, as originally presented in Fig. 2 of Ref.~\cite{Gluscevic:2010vv}. The dot-dashed lines show the threshold values of $r$ required to enable a $2,\,3\sigma$ detection of chirality for a model with $\Delta\chi=0.9$.}
\label{fig:sigmachi}
\end{figure}

We have calculated the CMB temperature and polarization anisotropy spectra for the example model. We have chosen non-inflationary cosmological parameters that are consistent with a best-fit $\Lambda$CDM model.  The distinguishing property of the CMB spectra in these scenarios is the prediction of non-zero parity-odd correlations $\langle TB\rangle$ and $\langle EB\rangle$. These spectra are illustrated in Fig.~\ref{fig:sigmachi}. The B-mode polarization power $\langle BB \rangle$ may soon be within reach of Stage-III CMB experiments, provided that foregrounds can be removed cleanly \cite{Abazajian:2016yjj}.

To assess the sensitivity of future experiments to parity-odd signals, we turn to Ref.~\cite{Gluscevic:2010vv}, where the $1\sigma$ error bars on $\Delta\chi$ are shown for future experiments. We reproduce the underlying calculations, shown in our Fig.~\ref{fig:sigmachi}, for an idealized satellite experiment (CMBpol) and a cosmic-variance limited experiment. If we consider a scenario with a chiral asymmetry $\Delta\chi=0.9$, as predicted in many of these models, then detection by a cosmic-variance limited experiment at the $2,\,3\sigma$ level would require a tensor-to-scalar ratio in excess of $r> 0.012,\, 0.027$. In this idealistic experimental situation, the $\nx=1/8$ model illustrated in the previous subsection would be close to the threshold of detection. We note, as pointed out in Ref.~\cite{Gluscevic:2010vv}, that the signal is dominated by low $\ell$ contributions to the temperature - B-mode correlation $\langle TB\rangle$, meaning a full sky experiment would be needed. In the case of a lower-amplitude signal or a less-than-idealized experiment, it would not be possible to discern the chiral pattern based on these correlations \cite{Gerbino:2016mqb}. In a sense, the problem lies in the two-dimensional nature of the gravitational wave imprint on the CMB. A 2D imprint cannot distinguish between left- and right-circular polarizations, but a 3D imprint can. Proposals to overcome this challenge using future galaxy- and 21-cm clustering surveys \cite{Jeong:2012df,Masui:2017fzw} might dramatically lower the threshold.
 
\subsection{Primordial Gravitational Wave Background}

\begin{figure}[h]
\includegraphics[width=0.65\linewidth]{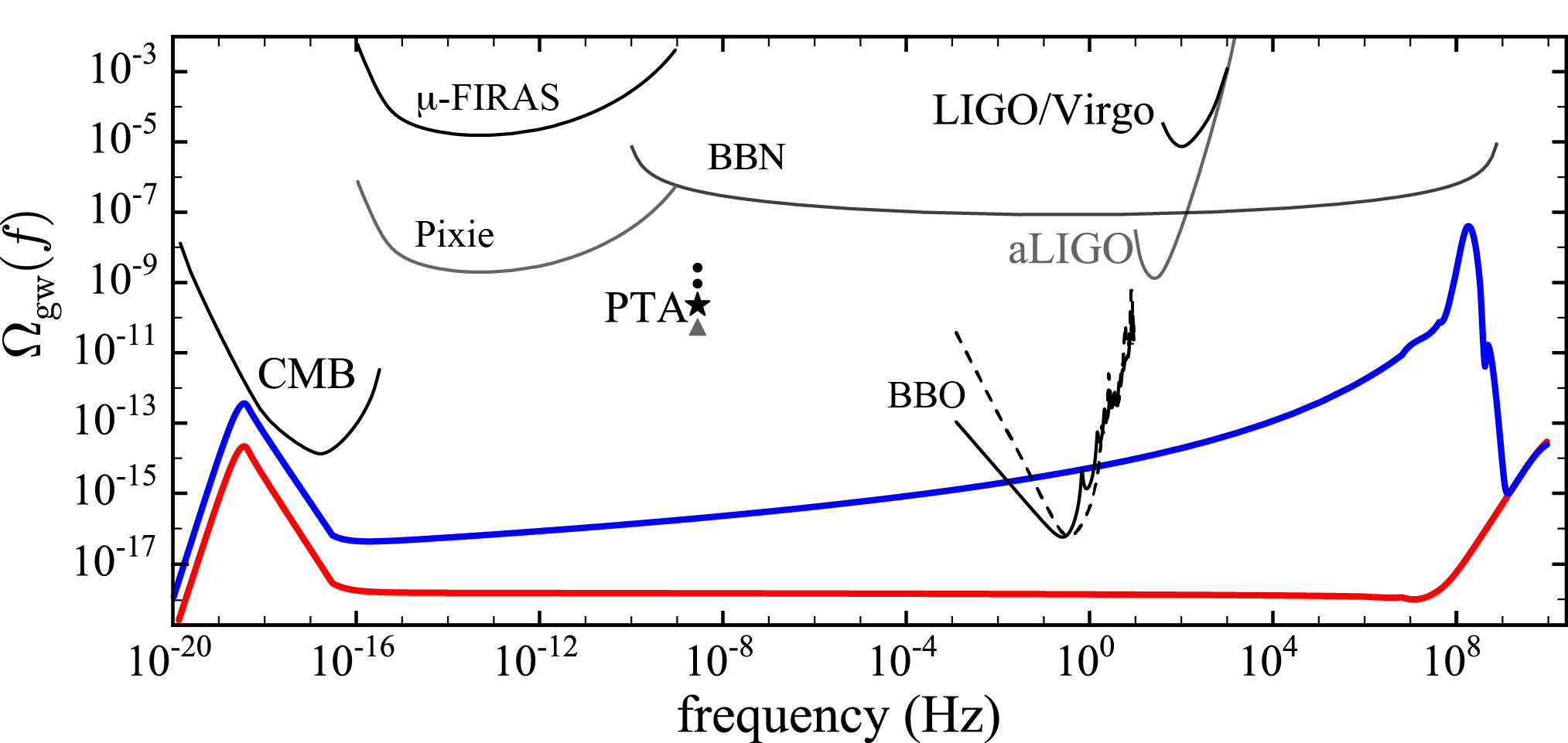}
\caption{The present-day gravitational wave spectral density $\Omega_{GW}$ is shown as a function of frequency for left- (blue) and right- (red) circularly polarized gravitational waves. The difference becomes highly pronounced near the end of inflation. However, the chiral asymmetry vanishes for modes deep inside the horizon, which is seen in the convergence at high frequency. The power-law integrated curves \cite{Thrane:2013oya} for the sensitivities of big bang nucleosynthesis (BBN), CMB (Planck), LIGO/Virgo, aLIGO, and Pulsar Timing Array (PTA) to a power-law stochastic background are adapted from Fig.~1 of Ref.~\cite{Lasky:2015lej}. For comparison, we show the projected sensitivity of BBO, a futuristic satellite-based gravitational wave observatory, to the intensity (solid) and circular polarization (dashed), which is adapted from Fig. 4 of Ref.~\cite{Smith:2016jqs}. The curve labeled $\mu-{\rm FIRAS}$ shows the bound on a stochastic gravitational wave background due to measurements of the degree of spectral distortion of the CMB, as calculated in Ref.~\cite{Chluba:2014qia}. The curve labeled ``Pixie" is a projection of the expected improvement on this bound by a futuristic, satellite-based spectral distortion experiment \cite{Kogut:2011xw}.}
\label{fig:gwspectrum}
\end{figure}

We have calculated the present-day spectral density of primordial gravitational waves predicted in this model of axionic gauge field inflation. To be definite, we follow the procedure given in Ref.~\cite{Watanabe:2006qe}, although we have not included the slight damping effects of neutrinos or the thermal history of the cosmological fluid. The spectrum we obtain is unique in two different ways. First, a blue-tilted spectrum has been achieved without violating the null energy condition (i.e. the expansion rate does not increase during inflation). The blue tilt means the spectrum might be within reach of future gravitational wave observatories. Second, the spectrum is dominated by left-circularly polarized gravitational waves which means that sensitivity to the V Stokes parameter of a stochastic gravitational wave background would be important to test this model. Recent work has shown that a pair of satellite gravitational wave observatories, with sensitivity beyond the reach of LISA \cite{AmaroSeoane:2012km,Crowder:2005nr}, would be required to detect the primordial V polarization \cite{Smith:2016jqs}. As illustrated in Fig.~\ref{fig:gwspectrum}, the predicted spectrum may be tested in two distinct frequency regimes, separated by nearly 17 orders of magnitude. Similar conclusions have recently been obtained by Ref.~\cite{Thorne:2017jft}. As we discuss next, there is a further, indirect test at the highest frequencies.
  
\section{Leptogenesis}

The chiral asymmetry of the gravitational wave background may help explain the matter-antimatter asymmetry of the Universe. This inflationary scenario contains the elements required for a leptogenesis scenario, similar to the scenario proposed in Ref.~\cite{Alexander:2004us}. The crux of the argument rests on index theorems which relate the properties of wave operators to a spacetime-curvature invariant \cite{Eguchi:1980jx}. In particular, the number of righthanded minus lefthanded axial vector solutions on a spacetime manifold is equal to $P/24$, where 
\begin{equation}
P =\frac{1}{16 \pi^2} \int d^4 x \sqrt{-g}\, R_{\mu\nu\alpha\beta} \widetilde R^{\mu\nu\alpha\beta}
\end{equation}
is the Pontryagin number \cite{Delbourgo:1972xb,Eguchi:1976db}, and
\begin{equation}
R \widetilde R \equiv R_{\mu\nu\alpha\beta} \widetilde R^{\mu\nu\alpha\beta} = \frac{1}{2} \epsilon_{\alpha\beta\mu\nu} R^{\alpha\beta\sigma\delta} R^{\mu\nu}{}_{\sigma\delta}.
\end{equation}
In physics, this result is expressed in the form of the gravitational anomaly for the lepton number current
\begin{equation}
\nabla_\mu J_\ell^\mu = \frac{N_{R-L}}{24(16 \pi^2)} R_{\mu\nu\alpha\beta} \widetilde R^{\mu\nu\alpha\beta}
\label{eqn:jrr}
\end{equation}
where $N_{R-L}$ is the number of righthanded minus lefthanded Weyl fermions \cite{AlvarezGaume:1983ig}. (We note that the factor of $24$ has been omitted from previous investigations \cite{Alexander:2004us,Maleknejad:2014wsa,Maleknejad:2016dci}.) Physically what happens is that the chirally asymmetric gravitational wave spectrum acts as a biased background for the evolution of the Dirac equation \cite{Hawking:1976jb}. Pairs of fermions are created, favoring one chirality over the other.  Coincidentally, this phenomenon has recently been observed in an analogue condensed matter system \cite{Gooth:2017mbd}.
  
In an idealized spacetime such as Bianchi IX, which resembles a single, circularly-polarized gravitational wave wrapped around a closed Robertson-Walker spacetime, this phenomenon imparts a handedness onto the spectrum of lepton creation \cite{Gibbons:1979ks}. In our scenario, the biased chirality originates at the perturbative level with the circularly-polarized gravitational wave background. Integration of the gravitational anomaly equation through the inflationary epoch shows that a lepton asymmetry is created. Leptogenesis in gauge field inflation models that use this process has been previously considered in Refs.~\cite{Maleknejad:2014wsa,Maleknejad:2016dci,Papageorgiou:2017yup}.

To determine the magnitude of the lepton asymmetry generated through this process, we define the number density of chiral fermions $n_\ell$ as determined by an observer with four-velocity $u$, by $n = -u \cdot J_\ell$. For a comoving, cosmological observer, then $n=a J_\ell^0$ where $n$ satisfies the differential equation
\begin{equation}
\frac{\partial}{\partial\tau} n_\ell + 3 \frac{a'}{a} n = \frac{N_{R-L}}{24(16 \pi^2)} a R\widetilde R.
\end{equation}
Recasting this as an integral, the solution is
\begin{equation}
n_\ell= \frac{N_{R-L}}{24(16 \pi^2) a^3} \int d\tau \, a^4 R \widetilde R. 
\label{eqn:RRdt}
\end{equation}
Next, we evaluate $R\widetilde R$ in our spacetime, to quadratic order in tensor gravitational wave perturbations. We obtain
\begin{eqnarray}
a^4 R\widetilde R &=&  \int \frac{d^3 k}{(2 \pi)^3}\frac{d^3 k'}{(2 \pi)^3} e^{i(\vec k + \vec k')\cdot \vec x} \left(I_R(\vec k,\,\vec k') - I_L(\vec k,\,\vec k')\right) \\
I_P &=& -4 \left( k'^2 k\, h_P(\vec k') h'_P(-\vec k) -k^2 k'\, h'_P(\vec k')h_P(-\vec k)  + k  h'_P(\vec k')h_P''(-\vec k) - k' h''_P(-\vec k) h'_P(\vec k') \right)
\end{eqnarray}
where $h_P$ are the Fourier amplitudes of polarization $P=R,\,L$. Next, we evaluate the expectation value $\langle n_\ell \rangle$: we convert the Fourier amplitudes to operators, and apply the commutation relations in the quantum state previously identified for the inflationary scenario. The expectation value is
\begin{eqnarray}
\langle n_\ell \rangle &=& \frac{N_{R-L}}{24(8 \pi^2) a^3} \int d\tau\, \int \frac{d^3 k}{(2 \pi)^3}\frac{d^3 k'}{(2 \pi)^3} e^{i(\vec k + \vec k')\cdot \vec x} (2 \pi)^3 \delta(\vec k + \vec k')  \left(F_R(\vec k) - F_L(\vec k)\right) 
\label{eqn:tderiv}\\
F_P &=& \frac{d}{d\tau}\left[ k^3 \left(  | h_{P,k}|^2_{{\mathscr H}_v} + | h_{P,k}|^2_{{\mathscr H}_u}  \right) -k \left(  | h'_{P,k}|^2_{{\mathscr H}_v} + | h'_{P,k}|^2_{{\mathscr H}_u}  \right)\right].
\end{eqnarray}
We now appreciate that the Chern-Pontryagin scalar is an exact divergence, as we convert the above expression into a boundary term. We assume that the difference between the right- and left-circularly polarized spectra vanished in the distant past (i.e. at the lower limit of integration in Eq.~(\ref{eqn:tderiv})), which is consistent with our inflationary model. Hence, the number density may be expressed in terms of the final gravitational wave spectra, 
\begin{eqnarray}
\langle n_\ell \rangle &=& \frac{N_{R-L}}{24(8\pi^2) a^3} \int d\ln k \, \left[ k^3 (\Delta_R^2 - \Delta_L^2) - k (\Delta_R^{\prime2} - \Delta_L^{\prime2}) \right] \\
\Delta_P  &=& \frac{k^3}{\pi^2}  \left(  | h_{P,k}|^2_{{\mathscr H}_v} + | h_{P,k}|^2_{{\mathscr H}_u}  \right), \qquad
\Delta_P^{\prime2} = \frac{k^3}{\pi^2}  \left(  | h'_{P,k}|^2_{{\mathscr H}_v} + | h'_{P,k}|^2_{{\mathscr H}_u}  \right),
\end{eqnarray}
which we evaluate at the end of inflation.

It is convenient to express the asymmetry in terms of a ratio between the lepton asymmetry density and the entropy density of the radiation fluid. This ratio is constant over the epoch of adiabatic evolution of the thermalized radiation fluid. We assume for convenience that reheating is instantaneous. That is, the energy density of the inflaton and gauge field at the end of inflation are converted into thermal radiation
\begin{equation}
\rho = 3 M_P^2 H_{end}^2 =  \frac{g_* \pi^2}{30}T^4
\end{equation}
with $g_*$ effective degrees of freedom. We solve for temperature, and express the entropy density $s = 2 g_{*s}\pi^2 T^3/45$ in terms of $H_{end}$, entropy degrees of freedom $g_{*s}$, and constants. The result is
\begin{equation}
\frac{\langle n_\ell \rangle}{s} =  \frac{N_{R-L}}{24(8 \pi^2) a_{end}^3} \frac{ \int d\ln k \, \left[ k^3 (\Delta_R^2 - \Delta_L^2) - k (\Delta_R^{\prime2} - \Delta_L^{\prime2}) \right]}{C g_*^{1/4} \left( H_{end} M_P \right)^{3/2}}
\label{eqn:nls}
\end{equation}
where $C = (128 \pi^2/45)^{1/4} \simeq 2.3$ and we set $g_{*s} = g_{*}$, again for simplicity. 

\begin{figure}[h]
\includegraphics[width=0.45\linewidth]{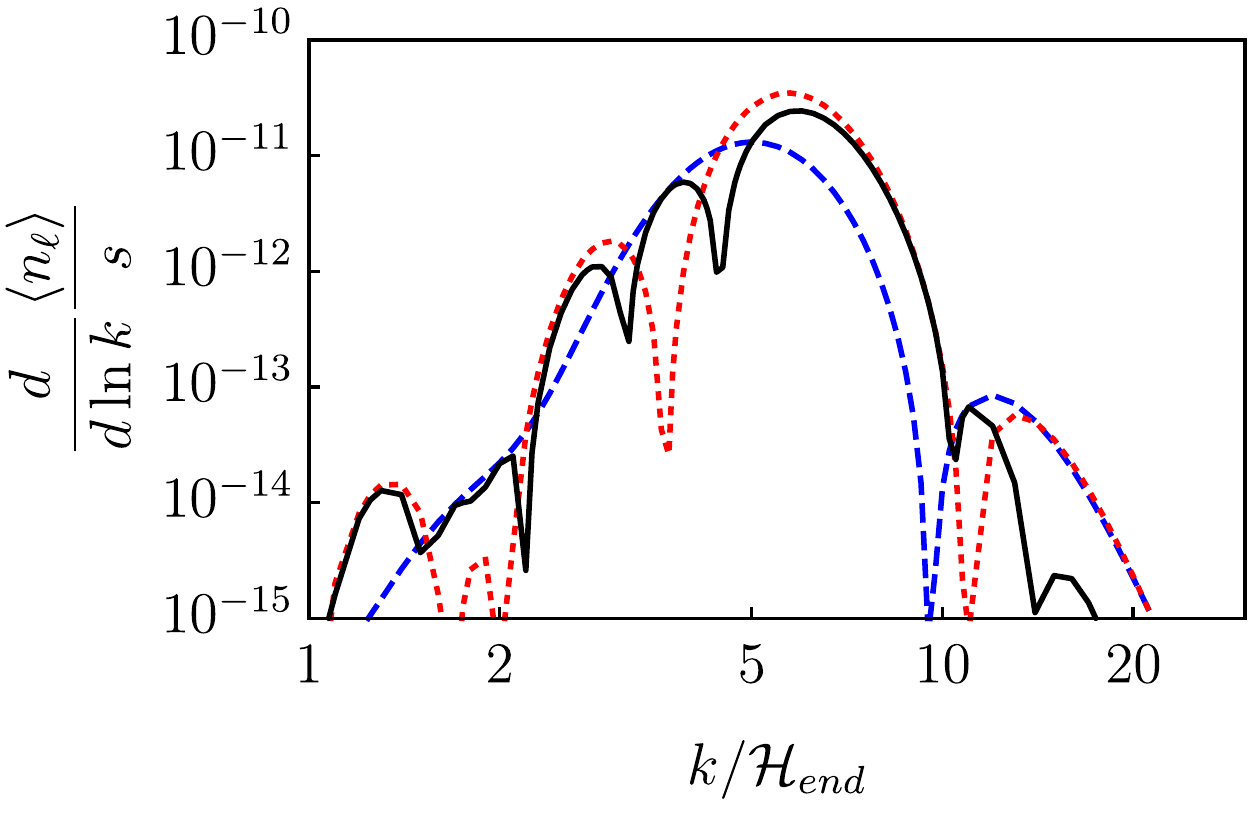}
\caption{
The integrand of Eq.~(\ref{eqn:nls}) is illustrated (solid, black) as a function of wavenumber, in units of the comoving Hubble scale at the end of inflation. Separate contributions due to the $k^3 (\Delta_R^2 - \Delta_L^2)$ (dashed, blue) and $k (\Delta_R^{\prime2} - \Delta_L^{\prime2})$ (dotted, red) terms are shown separately. In all cases the absolute value has been taken.}
\label{fig:lepto}
\end{figure}

We have evaluated $\langle n_\ell \rangle /s$ for a series of axionic gauge field inflation models. We first present results for the specific case of the model with $\nx=1/8$ and $r=0.035$, explored earlier in this paper. To consider a minimal model, we set $g_* = 106.75$ as for the Standard Model. We next fix $N_{R-L}=-3$ for the three left-handed Standard Model neutrinos. The integrand of Eq.~(\ref{eqn:nls}) is illustrated in Fig.~\ref{fig:lepto}, where the contributions from the $k^3$ and $k$ terms are shown separately. We obtain
\begin{equation}
\frac{\langle n_{\ell}\rangle}{s} = -2.45 \times 10^{-10}
\label{eqn:nlsnum}
\end{equation}
where the negative sign indicates a left-handed excess. 

The Sakharov conditions for successful baryogenesis are: violation of baryon number; CP violation; and the cosmic fluid should be out of equilibrium when these symmetry violations take place \cite{Sakharov:1967dj}.  This inflationary model satisfies these conditions by providing for the violation of lepton number through the gravitational anomaly, Eq.~(\ref{eqn:jrr}); the classical field configurations for the pseudoscalar axion $\chi$ and the gauge field $A_\mu^a$ are CP-asymmetric; the inflationary solutions for the vacuum expectation values of the fields are out of equilibrium. The conversion of a net lepton number into a net baryon number can occur through Standard Model electroweak sphaeleron processes. To determine the lepton-baryon exchange rate, we draw upon the result \cite{Kuzmin:1985mm,Khlebnikov:1988sr}
\begin{equation}
\Delta B = \frac{8 N_f + 4 N_H}{22 N_f + 13 N_H} \Delta L
\end{equation}
where $N_f$ is the number of lepton families and $N_H$ is the number of Higgs doublets. In the Standard Model, with three families and one Higgs doublet, this yields $\Delta B = \tfrac{28}{79} \Delta L$.

\begin{figure}[h]
\includegraphics[width=0.45\linewidth]{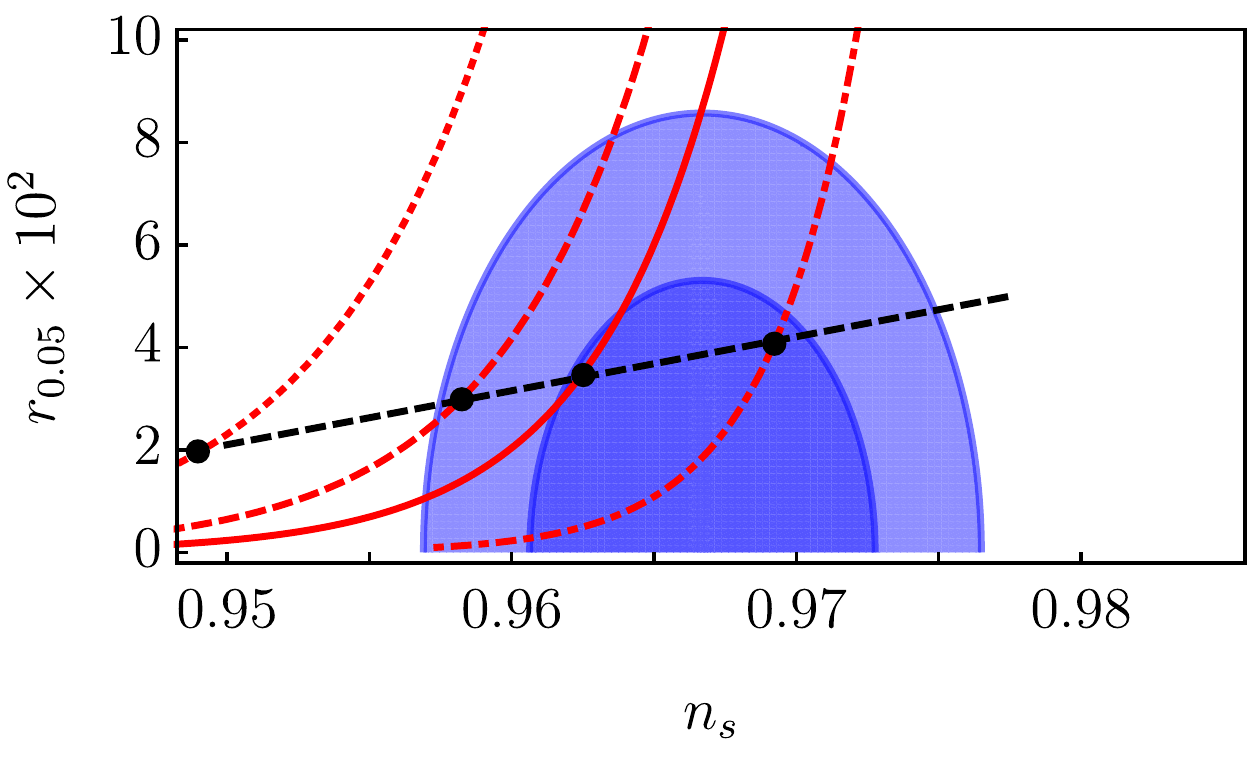}
\caption{Inflationary models with spectral index $n_s$ and tensor-to-scalar ratio $r_{0.05}$ that produce a baryon asymmetry $\eta = 6.1 \times 10^{-10}$ are indicated by the black circles for the cases of $\nx = 1/4$ (dotted, red), $1/6$ (dashed, red), $1/8$ (solid, red), and $1/16$ (dot-dashed, red). The dashed black line is a straight-line fit, showing that the tensor-to-scalar ratio for viable models that match the observed baryon asymmetry lies in the range $r_{0.05}\sim 3-4 \times 10^{-2}$.}
\label{fig:baryo}
\end{figure}

If the above assumptions are justified, then we may convert the above lepton excess into a baryon excess, whereby this model predicts
\begin{equation}
\eta \equiv \frac{n_B}{n_\gamma}= \frac{n_B}{s}/ \frac{n_\gamma}{s} = \left(-\frac{28}{79} \frac{\langle n_{\ell}\rangle}{s}\right)/\left(0.14\right) = 6.1 \times 10^{-10}.
\end{equation} 
We use $g_{*s}=3.91$ to calculate $n_{\ell}/{s}$, and we have inserted a minus sign to convert from left-handed leptons to baryons. This result matches the observed value $\eta = 6.10\,(\pm 0.04) \times 10^{-10}$ \cite{Ade:2015xua}. Hence, within the caveats of this toy model, the elements are in place for successful baryogenesis.

We have carried out this calculation for a range of inflationary models, varying $\nx$ and parameters $g$, $M$, and $m$. We have fixed $N_*=55$ and the scalar amplitude $\Delta_\zeta^2= 2.2 \times 10^{-9}$. For each value of the scalar field potential index $\nx$, the predicted value of $r$ lies along a curve in the $n_s - r$ plane, as shown in Fig.~\ref{fig:nsr}. Along each such curve, the predicted baryon to photon ratio is found to increase with increasing $r$. In Fig.~\ref{fig:baryo}  we have marked the point at which the predicted baryon to photon ratio matches the observed value with a black circle. We have fit these points to a straight line that runs across the viable range of $n_s$ with $r$ in the range $0.03-0.04$. Hence, if the model is to explain the baryon asymmetry of the Universe, then the predicted value of the tensor-to-scalar ratio lies in the range $0.03-0.04$. These models are a target for observation. Detection of a BB spectrum at the amplitude $r_{0.05}\sim 3-4 \times 10^{-2}$, with parity-violating TB and EB correlations, would be strongly suggestive of an axionic gauge field inflationary and leptogenesis scenario.

We have examined the spectrum of tensor gauge field waves as well, in order to confirm that they remain perturbative as do the gravitational waves. Recalling Eq.~(\ref{eqn:Atnsr}), we write the mean squared fluctuation amplitude as
\begin{equation}
\langle (\delta A)^2 \rangle \equiv \langle \delta A^a_\mu\, \delta A^b_\nu\rangle \, \delta_{ab} \, g^{\mu\nu} = \int\frac{d^3 k}{(2 \pi)^3}|u_k(\tau)|^2/a^2
\end{equation}
which is to be compared with $\langle A^2 \rangle = \langle A^a_\mu\, A^b_\nu \rangle \, \delta_{ab} \, g^{\mu\nu} =3 \phi^2/a^2$. On a mode by mode basis, we want to check that the ratio
\begin{equation}
\sigma_A^2(k) \equiv \frac{1}{\langle A^2 \rangle} \frac{d}{d \ln k}\langle (\delta A)^2 \rangle = \frac{k^3}{6 \pi^2 \phi^2} |u_k(\tau)|^2
\end{equation}
is less than unity. This criteria was also discussed for Higgsed gauge field inflationary models \cite{Adshead:2016omu,Adshead:2017hnc}. We have numerically evaluated $\sigma_A$ for a range of wavenumbers, in the case of several models that produce the observed baryon asymmetry, i.e. lying along the dashed line in Fig.~\ref{fig:baryo}. At the end of inflation, we find that $\sigma_A(k)$ is negligibly small for wavenumbers that lie outside the horizon. For the unamplified chirality, $\sigma_A$ is also tiny. However, $\sigma_A$ grows for left-circularly polarized modes that are just inside the horizon, similar to the peak seen in Fig.~\ref{fig:lepto}. We find that $\sigma^2_A \simeq 10^{-4}$ at its peak, $k/{\cal H}\simeq 5$, which suggests a root mean squared fluctuation $\delta A/A \lesssim 10^{-2}$ that appears to be safely perturbative.

We also calculate the spectral energy density
\begin{equation}
\Omega_{\delta A} = \frac{1}{\rho_{c}}\frac{d}{d \ln k} \langle \rho_{\delta A}\rangle 
\end{equation}
where the leading contribution to the fluctuation energy density at high frequency is $\rho_{\delta A} \simeq (t'^2+ k^2 t^2)/2a^2$ and $t = u/\sqrt{2} a$. The spectral energy density in the gauge field is more simply expressed as a fraction of the background gauge field energy density, whereby
\begin{equation}
\frac{1}{\rho_A} \frac{d}{d \ln k}  \langle \rho_{\delta A}\rangle = \frac{2}{3 (\phi'^2 + g^2 \phi^4)}\frac{k^2}{8 \pi^2}\left(  |u'_k(\tau)|^2 + k^2  |u_k(\tau)|^2\right).
\end{equation}
Following a numerical calculation similar to that described in the above paragraph, we find that $\Omega_{\delta A} \ll 1$ at the end of inflation, peaking at an amplitude $\lesssim 10^{-4}$ for left-circularly polarized modes that are just inside the horizon, with $k/{\cal H}\simeq 5$. From this we conclude that the energy in the gauge field remains safely in the perturbative regime. We note that this amplitude exceeds the nucleosynthesis bound illustrated in Fig.~\ref{fig:gwspectrum}. However, we do not expect these tensor waves in the gauge field to survive through the nucleosynthesis era. Rather, these subhorizon tensor waves will participate in the reheating process and, with the gauge field itself, convert into Standard Model radiation. For tensor waves that are well outside the horizon at the end of inflation, the nucleosynthesis \cite{Cyburt:2015mya} and cosmic microwave background bounds \cite{Smith:2006nka} on energy density are easily satisfied in these models.
  
Here we discuss the parameter dependences. For a given potential exponent $\nx$, due to a degeneracy among the three parameters ($g$, $M$, $m$), models with a normalized scalar amplitude lie along a line as shown back in Fig.~\ref{fig:nsr}. Along such a line, tracing an upwards path of increasing $r$, the baryon excess $\eta$ grows nonlinearly. For example, in the vicinity of $\eta = 6.1 \times 10^{-10}$, $\eta$ scales as $\eta \propto r^{8/3}$.

We have assumed the inflaton and gauge field instantaneously thermalize into a relativistic bath of $g_*= 106.75$ degrees of freedom. If we allow the number of degrees of freedom to increase, which is entirely reasonable, then the lepton excess $\langle n_\ell\rangle/s$ will decrease. To achieve the observed value of $\eta$ then we will have to raise $r$. Hence, our current estimate of the value of $r$ that gives the observed $\eta$ is in fact a lower bound.

We have fixed the reference wavenumber $k = 0.05$~inv-Mpc to correspond to modes that depart the horizon $N_*=55$ e-foldings before the end of inflation.  In other models of inflation, the number $N_*$ may range from approximately $50-60$. We do not have this same freedom in this model, as discussed earlier.  However, suppose we were to assume that thermalization is not instantaneous, but is delayed by $N_{therm}$ e-foldings at the end of inflation. Depending on the equation of state of the dominant form of energy during this pre-thermalization stage, the predicted ratio $\langle n_\ell\rangle/s$ would shift up or down. In most inflationary models, the inflaton oscillates at the bottom of its potential at the end of the accelerated expansion, yielding a matter-dominated pre-thermalization epoch. In our case, however, the gauge field dominates at the end of inflation, so that the equation of state is rapidly driven to $w=1/3$. In this case, a delay in thermalization makes no change in our prediction of $\langle n_\ell\rangle/s$ and therefore $\eta$.

We note that the integration is dominated by the highest frequency modes, meaning those modes that are still subhorizon or have just exited the horizon at the end of inflation. Here we make substantial improvement relative to previous estimates of the degree of lepton asymmetry generated through chiral primordial gravitational waves. Whereas previous investigations simply cut off the integration at the wavenumber corresponding to the horizon radius at the end of inflation, we continue our integration to slightly higher wavenumbers corresponding to subhorizon modes. Our cutoff is provided naturally, since the asymmetry between left- and right-circular polarizations drops off rapidly for subhorizon modes. Or to put it another way, there is no chiral asymmetry for deep subhorizon modes; instead, the equations of motion distinguish the handedness as modes begin to approach horizon crossing. The drop off in chirality can be seen in the behavior of the integrands in Fig.~\ref{fig:lepto} as well as in the convergence of left- and right-hands of the tail of the gravitational wave spectrum in Fig.~\ref{fig:gwspectrum}. 

Finally, we point out that if right handed neutrinos are produced in reheating, then the slight lepton asymmetry created through gravitional processes can be erased. Therefore, to preserve the lepton asymmetry, we must assume the neutrino mass $m_{\nu R}$ is much greater than the reheat temperature, $T_{rh}$. If thermalization is instantaneous, then $T_{rh} =(90 M_P^2 H_{end}^2/g_* \pi^2)^{1/4} \simeq 3\times 10^{15}$~GeV, where $H_{end} = 4 \times 10^{-6} M_P$ for the model in question, which sets a high threshold for the neutrino mass. This is roughly consistent with the heavy neutrino masses obtained by a see-saw mechanism \cite{Mohapatra:1979ia,King:2003jb}. If thermalization is delayed, leading to a lower reheat temperature, then the bound on the neutrino mass would be similarly reduced. We also note that if right handed neutrinos exist, then we implicitly assume their masses are above the energy scale corresponding to the gravitational wave frequencies that dominate the integrals in Eq.~(\ref{eqn:nls}). For the model in question the masses must exceed $m_{\nu_R} \gg 5 \times H_{end} \simeq 5 \times 10^{13}$~GeV to be safely out of reach. Clearly, advancing this scenario beyond that of a toy model will require a more complete treatment of reheating.

\section{Summary}

In this work we have presented a toy model of inflation that features an inflaton and a gauge field. The inflaton potential itself is far too steep to yield slow roll inflation on its own. However, the coupling between the gauge field and the inflaton effectively flattens the potential, and inflation ensues. All mass scales and the field excursion during inflation are well below the Planck scale, and all dimensionless couplings are small but not finely tuned. The epoch ends naturally when the inflaton reaches the bottom of its potential, whereupon the gauge field dominates with equation of state $w=1/3$. We have shown that density and gravitational wave spectra are produced with amplitude and spectral tilt which are consistent with current observations, as illustrated in Fig.~\ref{fig:nsr}. 

The most remarkable features of the spectrum of gravitational waves are the circular polarization and the blue tilt that extends out to high frequencies. We have explored the consequences of this chiral asymmetry for the polarization pattern imprinted on the cosmic microwave background and for the direct detection of the stochastic gravitational wave background by a future, satellite-based interferometric observatory. The circular polarization signal is shown to be within reach of both means of detection, and offers a distinct method to test this scenario. Finally, the blue tilt and chiral asymmetry provide key ingredients for a leptogenesis scenario to explain the matter-antimatter asymmetry of the Universe. If this toy inflationary model is to explain the observed baryon asymmetry, then we predict a tensor-to-scalar ratio of no less than $r_{0.05} \sim 3-4\times 10^{-2}$, as shown in Fig.~\ref{fig:baryo}.

We have explored many aspects of this inflationary model, but many more investigations lie ahead. Our method of numerical calculation of the spectra is inefficient, and so we have kept a narrow focus and made simplifying assumptions.  Within the confines of this toy model, we have yet to explore the full parameter predictions. We leave for future work the study of non-gaussianity and bispectra \cite{Agrawal:2017awz}; the behavior of vector modes; and a more realistic treatment of the particle physics background of this model.

\acknowledgments
This work is supported in part by DOE grant DE-SC0010386. We thank Peter Adshead, Tristan Smith, Marco Peloso and Alex Papageorgiou for useful discussions. We thank Nordita for bringing together many of the experts on cosmological gauge fields to the July 2017 workshop on Inflation and the CMB.


\appendix
\section{SU(N)}
\label{app:SUN}

We have investigated inflationary scenarios in which the axionic inflaton is coupled to a gauge field that is symmetric under a larger group, SU(N). To maintain homogeneity and isotropy of the field energy, we extend our flavor-space locked configuration to the ${\cal N} = [N/2]$ disjoint SU(2) subgroups within an SU(N). Hence, two such subgroups can be embedded in SU(4) and SU(5), three in SU(6), SU(7), etc. The field strength tensor is
\be
	F^{a}_\mn \equiv \partial_\mu A^{a}_\nu - \partial_\nu A^{a}_\mu -g f^{abc}A_{b\mu} A_{c\nu},
\ee
where $f^{abc}$ are the structure constants for the relevant gauge group. The vector field $A^b_\mu$ equals a different scalar $\phi_n$ in each subgroup, where $n$ labels the subgroup. The energy density and pressure are
\begin{equation}
\rho =\frac{3}{2 a^4}\sum_{n=1}^{\cal N} \left(\phi_n'^2 + g^2 \phi_n^4\right) + \frac{1}{2}\left(\frac{\chi'}{a}\right)^2 + V, \qquad
p = \frac{1}{2 a^4}\sum_{n=1}^{\cal N} \left(\phi_n'^2 + g^2 \phi_n^4\right)  + \frac{1}{2}\left(\frac{\chi'}{a}\right)^2 - V.
\end{equation}
The equations of motion are
\begin{equation}
 \chi'' + 2 \frac{a'}{a}\chi' + a^2 V_{,\chi} = 12 \frac{g}{a^2 M}\sum_{n=1}^{\cal N}  \phi_n^2\phi_n' , \qquad
\phi_n'' + 2g^2 \phi_n^3 + 4 g \phi_n^2 \frac{\chi'}{M}=0.
\end{equation}
The phase space for ${\cal N}>1$ rapidly becomes difficult to track. However, our numerical experimentation reveals two simplifying results. First, when there is more than one subgroup, a subset of the gauge fields will dominate. These dominant fields rapidly evolve towards a common field strength $\phi$, and guide the scalar $\chi$ onto the accelerating track.  Second, the remaining gauge fields that are subdominant will dilute away like radiation. The resulting picture of axionic gauge field inflation with multiple subgroups is that there will be a single field strength $\phi$ for each of the ${\cal N}_d$ dominant subgroups. The energy density and pressure become
\begin{equation}
\rho = \frac{3 {\cal N}_d}{2 a^4}  \left(\phi'^2 + g^2 \phi^4\right) + \frac{1}{2}\left(\frac{\chi'}{a}\right)^2 + V, \qquad
p = \frac{ {\cal N}_d}{2 a^4} \left(\phi'^2 + g^2 \phi^4\right)  + \frac{1}{2}\left(\frac{\chi'}{a}\right)^2 - V.
\end{equation}
The equations of motion are
\begin{equation}
\chi'' + 2 \frac{a'}{a}\chi' + a^2 V_{,\chi} = 12  {\cal N}_d \frac{g}{a^2 M} \phi^2\phi' , \qquad
\phi'' + 2g^2 \phi^3 + 4 g \phi^2 \frac{\chi'}{M}=0.
\end{equation}
There is a further simplification. By replacing $\phi \to \phi/\sqrt{{\cal N}_d}$ and $g \to g \sqrt{{\cal N}_d}$ all background equations can be made equivalent to the original case with a single SU(2). It is not obvious whether this scaling can bring the fluctuation spectrum into agreement with SU(2), too, since there are still ${\cal N}_d$ fluctuating fields.

To evaluate the scalar perturbation spectra, we englarge  Eq.~(\ref{eqn:dAtnsr}) to allow for scalar perturbations in each of the ${\cal N}$ subgroups. We express the second order action in terms of the dynamical degrees of freedom $X = \{\delta M_1,\, \delta Q_1, \delta M_2,\, \delta Q_2, ...\, \delta M_{\cal N},\, \delta Q_{\cal N},\,\delta\chi \}$ and $N=\{Y_1,\, Y_2,\, ... \,Y_{\cal N},\, B,\, \Phi\}$ are the constraints.  Hence there are $2{\cal N}+1$ degrees of freedom and ${\cal N}+2$ constraints. Operationally, the procedure follows the case outlined for SU(2) in Sec.~\ref{sec:scalar}.

The gravitational wave equations for an axion gauge field inflation scenario with ${\cal N}$ subgroups are as follows:
\begin{eqnarray}
&&v_L'' + \left[k^2 - \frac{a''}{a}  + \frac{2{\cal N}}{a^2 M_P^2}(g^2 \phi^4 - \phi'^2)\right] v_L = \frac{2}{a M_P}\sum_{n=1}^{\cal N} \left[ (g \phi + k) g  \phi^2 u_{Ln} -  \phi' u_{Ln}'\right]
\label{eqn:ivL} \\
&&u_{Ln}'' + \left[k^2 + 2 g k \phi +4(g\phi+k)\frac{\chi'}{M} \right] u_{Ln}  
=\frac{2}{a M_P}\left[ a\left(\frac{v_L}{a} \right)'\phi' + g \phi^2 \left(k-g\phi +4 \frac{\chi'}{M}\right)  v_L  \right].
\label{eqn:iuL}
\end{eqnarray}
The equations for $v_R,\,u_{Rn}$ are obtained by replacing $k \to -k$. To evolve the quantum fluctuations in $v_L$ requires ${\cal N}+1$ sets of Eqs.~(\ref{eqn:ivL}-\ref{eqn:iuL}). For the first set, the initial conditions are
\begin{equation}
v_L |_i = \frac{e^{-i k \tau_i}}{\sqrt{2 k}}, \quad v'_L |_i =-i k \frac{e^{-i k \tau_i}}{\sqrt{2 k}}, \quad u_{Ln} |_i = u'_{Ln} |_i = 0.
\end{equation}
We will refer to the solution to $v_L$ for this set of initial conditions as ``$v_{LH}$'' where the $H$ indicates the homogeneous solution.
For each subsequent set, for $n=1,\,2,\, ... \, {\cal N}$, the initial conditions are
\begin{equation}
v_L |_i = v'_L |_i =0, \quad u_{Ln} |_i = \frac{e^{-i k \tau_i}}{\sqrt{2 k}}, \quad u'_{Ln} |_i = -i k \frac{e^{-i k \tau_i}}{\sqrt{2 k}}.
\end{equation}
We will refer to the solutions to $v_L$ for these sets of initial conditions as ``$v_{LIn}$'' for the $n-$th inhomogeneous solution. Each of these inhomogeneous solutions are the same, so we will drop the ``$n$" from the subscript. The power spectrum is obtained by adding in quadrature the homogeneous solution for $v_L$ with the ${\cal N}$ inhomogeneous solutions for $v_L$ due to each $u_{Ln}$. Hence
\begin{eqnarray}
\Delta_L^2 &=& \frac{k^3}{2 \pi^2}\frac{4}{a^2 M_P^2}\left( |v_{LH}|^2 + \sum_{n=1}^{\cal N} |v_{LIn}|^2 \right) \cr
&=& \frac{k^3}{2 \pi^2}\frac{4}{a^2 M_P^2}\left( |v_{LH}|^2 +  {\cal N} |v_{LI}|^2 \right)
\end{eqnarray}
gives the left-circularly polarized gravitational wave power spectrum. 

We have evaluated the scalar and tensor spectra for ${\cal N}=2$ for the specific case of $\nx=2$ in the gauge-flation picture. Gauge-flation \cite{Maleknejad:2011jw} is a twin model of inflation to chromo-natural inflation when the scalar field is on the accelerating track \cite{Adshead:2012qe}. Formally, the action for the theory can be obtained from Eq.~(\ref{eqn:thy}) by integrating out the scalar $\chi$. In the scalar sector, there are $2{\cal N}$ degrees of freedom, however, the two dominant modes are the same as in the ${\cal N}=1$ case. As a consequence, the parameters $g,\,m,\,M$ can be chosen to bring the ${\cal N}=2$ scalar spectrum into agreement with the ${\cal N}=1$ case. We have also made analytic calculations that suggest this agreement occurs for higher ${\cal N}$. For the tensors, the same rescaling $\phi \to \phi/\sqrt{{\cal N}}$ can be extended to the gauge field waves, and thereby bring the ${\cal N}=2$ tensor spectrum into agreement with the ${\cal N}=1$ case. Hence, both the ${\cal N}=2$ scalar and tensor spectra are identical to the ${\cal N}=1$ case.



\end{document}